\documentclass[aps,prb,showpacs,preprintnumbers,amsmath,amssymb,twocolumn,letterpaper,unsortedaddress]{revtex4}
\usepackage{amssymb}
\usepackage{ifpdf}
\usepackage[utf8]{inputenc}
\usepackage{graphicx}
\usepackage{hyperref}
\usepackage{bm}
\usepackage{textcomp}
\usepackage{times}

\newcommand{\LaThOBiS}{La$_{1-x}$Th$_{x}$OBiS$_2$}
\newcommand{\LaHfOBiS}{La$_{1-x}$Hf$_{x}$OBiS$_2$}
\newcommand{\LaTiOBiS}{La$_{1-x}$Ti$_{x}$OBiS$_2$}
\newcommand{\LaZrOBiS}{La$_{1-x}$Zr$_{x}$OBiS$_2$}
\newcommand{\LaSrOBiS}{La$_{1-x}$Sr$_{x}$OBiS$_2$}
\newcommand{\LaOBiS}{LaOBiS$_2$}

\newcommand{\ThOBiS}{ThOBiS$_2$}

\begin{document}
\title{Superconductivity induced by electron doping in the system La$_{1-x}$\emph{M}$_x$OBiS$_2$ (\emph{M} = Ti, Zr, Hf, Th) }

\author{D. Yazici$^1$}
\author{K. Huang$^1$}
\author{B. D. White$^1$}
\author{I. Jeon$^1$}
\author{V. W. Burnett$^1$}
\author{A. J. Friedman$^1$}
\author{I. K. Lum$^1$}
\author{M. Nallaiyan$^2$}
\author{S. Spagna$^2$}
\author{M. B. Maple$^1$}
\email[Corresponding Author: ]{mbmaple@ucsd.edu}
\affiliation{$^1$Department of Physics, University of California, San Diego, La Jolla, California 92093, USA}
\affiliation{$^2$Quantum Design, 6325 Lusk Boulevard, San Diego, California 92121, USA}
\date{\today}

\begin{abstract}
We report a strategy to induce superconductivity in the BiS$_2$-based compound LaOBiS$_2$. Instead of substituting F for O, we increase the charge-carrier density (electron dope) via substitution of tetravalent Th$^{+4}$, Hf$^{+4}$, Zr$^{+4}$, and Ti$^{+4}$ for trivalent La$^{+3}$. It is found that both the LaOBiS$_2$ and ThOBiS$_2$ parent compounds are bad metals and that superconductivity is induced by electron doping with \emph{T$_c$} values of up to 2.85 K. The superconducting and normal states were characterized by electrical resistivity, magnetic susceptibility, and heat capacity measurements. We also demonstrate that reducing the charge-carrier density (hole doping) via substitution of divalent Sr$^{+2}$ for La$^{+3}$ does not induce superconductivity.

\end{abstract}

% maximum 600 characters including spaces!
% PACS, the Physics and Astronomy Classification Scheme.
%74.70.Dd: Superconducting materials, multinary compounds
%74.25.Op: superconductors critical field
%74.25.F-: Superconductors, transport properties see 75.30.Mb: see also 71.27.+a Strongly correlated electron systems, heavy fermions)
\vskip2pc

\maketitle

\section{Introduction}

Superconductivity with $T_c$ = 8.6 K has recently been reported in the layered compound Bi$_4$O$_4$S$_3$. \cite{Mizuguchi1} Following this report, several other BiS$_2$-based superconductors, including $Ln$O$_{1-x}$F$_x$BiS$_2$ (\emph{Ln} = La, Ce, Pr, Nd, Yb) with $T_c$ as high as 10 K, have been synthesized and studied. \cite{Li, Jha, Deguchi, Kotegawa, Awana, Demura, Mizuguchi2, Xing, Yazici} These materials have a layered crystal structure composed of superconducting BiS$_2$ layers and blocking layers of Bi$_4$O$_4$(SO$_4$)$_{1-x}$ for Bi$_4$O$_4$S$_3$ and $Ln$O for $Ln$O$_{1-x}$F$_x$BiS$_2$ (\emph{Ln} = La, Ce, Pr, Nd and Yb). This structural configuration is similar to the situation encountered in the high-$T_c$ layered cuprate and Fe-pnictide superconductors, in which superconductivity primarily resides in CuO$_2$ planes and Fe-pnictide layers, respectively.\cite{Zhano, Alff, Paglione, Johnston, Mazin} Even though BiS$_2$-based superconductors share a similar crystal structure with Fe-pnictide superconductors, they exhibit some important differences. The undoped parent compounds, \emph{Ln}FeAsO, display a spin density wave (SDW) or a structural instability near 150 K.\cite{Paglione, Johnston, Mazin} ~Superconductivity emerges when the SDW is suppressed towards zero temperature either through charge carrier doping or application of pressure.\cite{okada_2008_1} $T_c$ is raised as high
as 55 K by replacement of La by other rare-earth element, Sm.\cite{Paglione, Johnston} ~In contrast, the phosphorus-based analogues \emph{Ln}FePO do not show a SDW transition or structural instability but still exhibit superconductivity.\cite{Hamlin, Kamihara} For LaFePO, values of $T_c$ that range from 3 K \cite{Kamihara} to 7 K, \cite{tegel_2008_1} without charge carrier doping, have been reported. Both \emph{Ln}FeAsO and \emph{Ln}FePO remain metallic to low temperatures, while the undoped parent compounds \emph{Ln}OBiS$_2$ are bad metals. It has been suggested that superconductivity emerges in close proximity to an insulating normal state for the optimal superconducting sample.\cite{Xing}

Several distinct examples of chemical substitution have been found to induce superconductivity in \emph{Ln}FeAsO compounds including substituting F for O, \cite{kamihara_2008_1, chen_2008_1, ren_2008_1, ren_2008_3, ren_2008_2, fang_2008_2, Xing} Co for Fe, \cite{sales_2008_2} Sr for La, \cite{zhu_2008_1} Th for Gd, \cite{ren_2008_5} and also the introduction of oxygen vacancies. \cite{ren_2008_6, yang_2008_1} Substituting F for O induces superconductivity in BiS$_2$-based superconductors.\cite{Yildirim, Li, Deguchi, Jha, Kotegawa, Awana, Demura, Mizuguchi2, Lee, Shein, Wan, Xing, Yazici} To determine whether superconductivity might emerge under other conditions, we chose to dope electrons via chemical substitution on the La site in LaOBiS$_2$.
In this study, we demonstrate that substitution of tetravalent Th$^{+4}$, Hf$^{+4}$, Zr$^{+4}$, and Ti$^{+4}$ for trivalent La$^{+3}$ in \LaOBiS\ induces superconductivity. We also observed that substitution of divalent Sr$^{+2}$ for La$^{+3}$ (hole doping) does not induce superconductivity.

\section{Experimental Methods}
Polycrystalline samples of \LaThOBiS~ ($0 \leq x \leq 1$), \LaHfOBiS~ ($0 \leq x \leq 0.4$), \LaZrOBiS~ ($0 \leq x \leq 0.3$), \LaTiOBiS~ ($0 \leq x \leq 0.3$), and \LaSrOBiS~ ($0 \leq x \leq 0.3$) were prepared by a two-step solid state reaction method using high-purity starting materials. Initially, the Bi$_2$S$_3$ precursor powders were prepared by reacting Bi and S grains together at 500 \textcelsius{} in an evacuated quartz tube for 10 hours. Starting materials of La, La$_2$O$_3$, S and  Bi$_2$S$_3$ powders and either \emph{M} = Th or Ti chunks, Hf granules, or Zr foil, were weighed in stoichiometric ratios based on nominal concentrations La$_{1-x}M_x$BiS$_2$ (\emph{M} = Th, Hf, Zr, Ti). Next they were thoroughly mixed, pressed into pellets, sealed in evacuated quartz tubes, and annealed at 865 \textcelsius{} for 72 hours.  Additional regrinding and sintering at 700 \textcelsius{} for 3 days was performed to promote phase homogeneity.  The crystal structure was verified by means of x-ray powder diffraction (XRD) using a Bruker D8 Discover x-ray diffractometer with Cu-K$\alpha$ radiation. The resulting XRD patterns were fitted via Rietveld refinement\cite{Rietveld} using the GSAS+EXPGUI software package.\cite{LARSON01,expgui} Electrical resistivity measurements were performed using a home-built probe in a liquid $^4$He Dewar for temperatures 1~K~$\leq$~$T$~$\leq$ 300~K by means of a standard four-wire technique using a Linear Research LR700 AC resistance bridge. Magnetization measurements were made for 2~K~$\leq$~$T$~$\leq$ 300~K and in magnetic fields $H$ = 5 Oe using a Quantum Design MPMS. AC magnetic susceptibility was measured down to $\sim$ 1 K in a liquid $^4$He Dewar. Specific heat measurements and electrical resistivity measurements under applied magnetic field for La$_{0.85}$Th$_{0.15}$OBiS$_2$ and La$_{0.8}$Hf$_{0.2}$OBiS$_2$ samples were performed for 0.36~K~$\leq$~$T$~$\leq$ 30~K in a Quantum Design PPMS DynaCool with a $^3$He insert using a thermal relaxation technique.

\section{Results}

\subsection{Crystal structure and sample quality}

\begin{figure}[t]
  \includegraphics[width=0.99\linewidth]{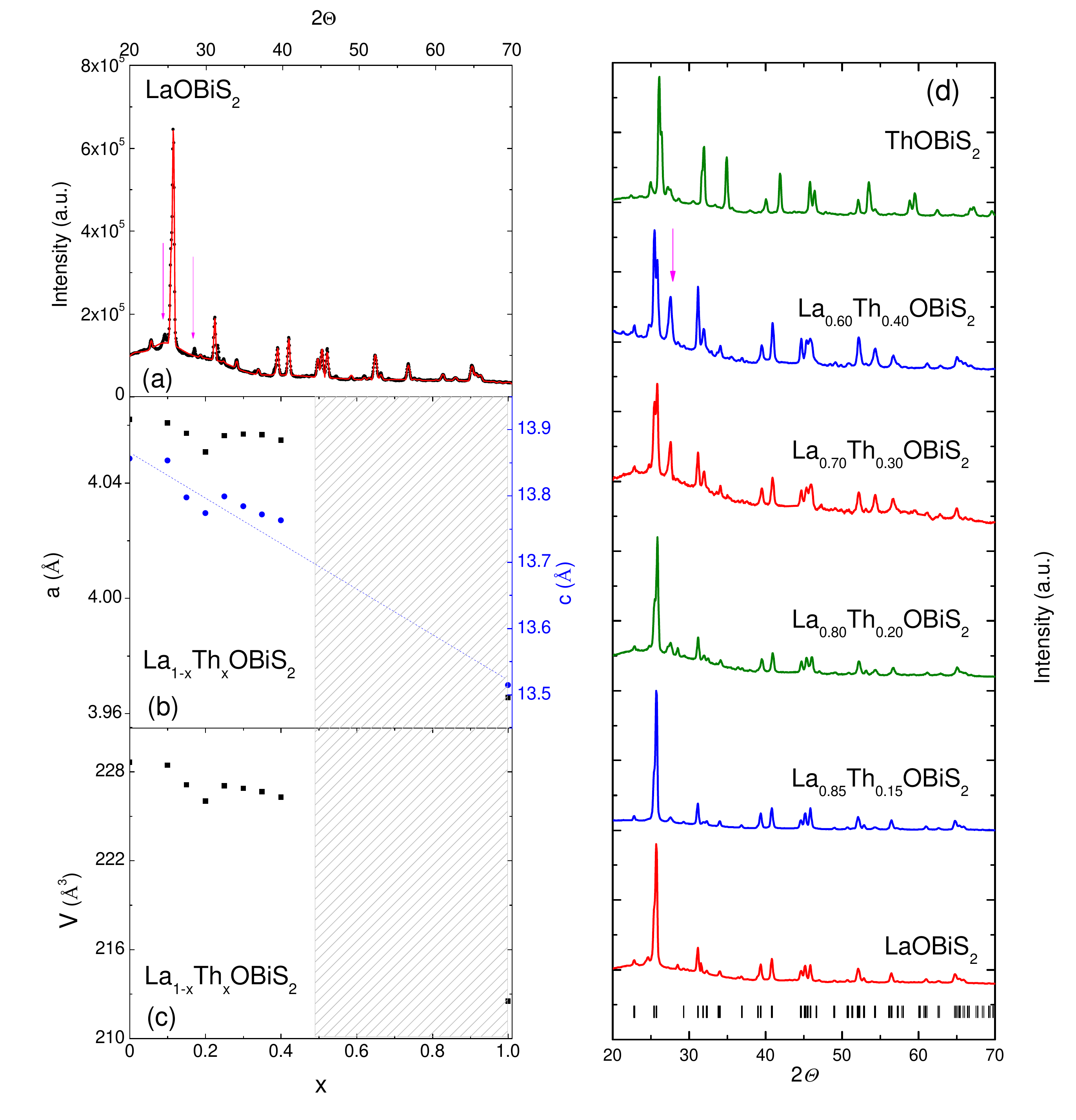}
  \caption{(Color online) (a) X-ray diffraction (XRD) pattern for LaOBiS$_2$. Black circles represent data, and the red solid line is the result of Rietveld refinement of the data. (b) Lattice parameters $a$ and $c$ vs. nominal Th concentration $x$. The gray region indicates a solubility limit near x = 0.45. (c) Unit cell volume $V$ vs. nominal Th concentration $x$. (d) XRD patterns for selected concentrations of \LaThOBiS\, where values of $x$ are explicitly labeled.}\label{fig:xrd}
\end{figure}

\begin{figure}[b!]
  \includegraphics[width=0.99\linewidth]{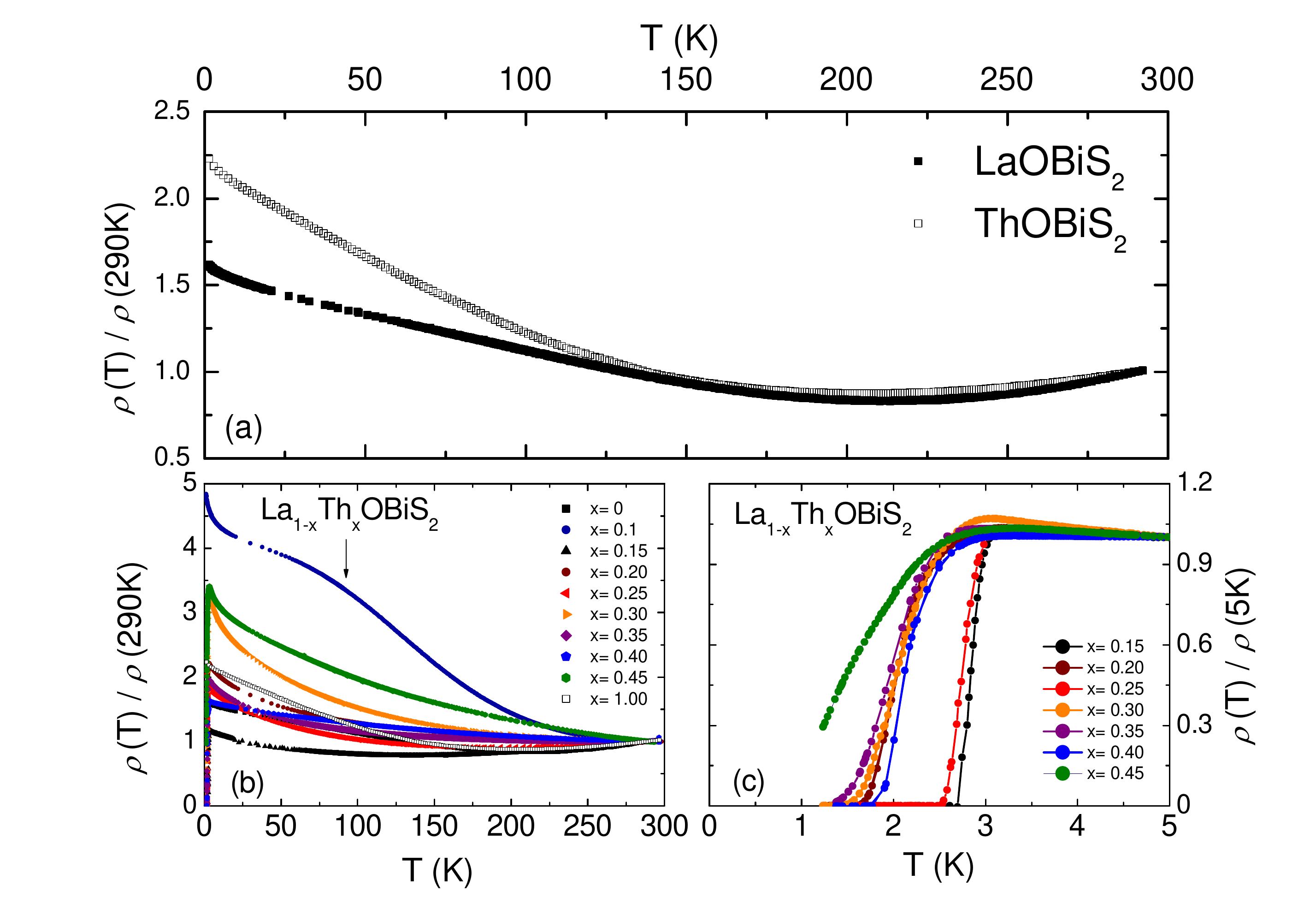}
  \caption{(Color online) (a) Electrical resistivity $\rho$, normalized by its value at $290$ K, vs.~temperature $T$ for \LaOBiS and \ThOBiS. (b) $\rho$, normalized by its value at $290$ K, vs. $T$ for \LaThOBiS~ ($0 \leq x \leq 1$). (c) $\rho$, normalized to its value in the normal state at 5 K, vs. $T$ emphasizing the transition into the superconducting state.}\label{fig:rho-Th}
\end{figure}

\LaOBiS\ and \ThOBiS\ are isostructural with the ZrCuSiAs structure type (space group $P4/nmm$). The inset of Fig.~\ref{fig:xrd}(a), displays the crystal structure of LaOBiS$_2$. The structure is composed of stacked La$_2$O$_2$ layers and two BiS$_2$ layers in each unit cell. To dope electrons into the BiS$_2$ conduction layers, we substituted tetravalent Th$^{+4}$, Hf$^{+4}$, Zr$^{+4}$, and Ti$^{+4}$ for La$^{+3}$ and to dope holes into the BiS$_2$ conduction layers, we substituted divalent Sr$^{+2}$ for  trivalent La$^{+3}$. We find that samples for the entire range of Th substitutions below the solubility limit, Hf substitutions in the range $0.1 \leq x \leq 0.4$, and Zr, Ti, and Sr substitutions in the range $0.1 \leq x \leq 0.3$, can be described by the same space group. As shown in Fig.~\ref{fig:xrd}(b), the lattice parameters $a$ and $c$ for \LaThOBiS~ ($0 \leq x \leq 1$) decrease with increasing Th concentration, although the relative decrease of $a$ is much smaller than that of $c$. With Th doping, the positions of Bragg reflections shift systematically. For example, the (004) reflections shift to higher angles, consistent with a decrease of $c$. This result indicates that Th is really incorporated into the lattice; although, small amounts of elemental Bi, indicated by arrows in Fig.~\ref{fig:xrd}(a, d), separate from the lattice. The amount of Bi impurity increases with increasing Th concentration. For $0.5 \leq x \leq 0.9$, the lattice parameters (not shown) are close to the lattice parameters of \ThOBiS\ and do not change appreciably with $x$. The Bi impurity amount increases with $x$ and other impurity peaks begin to appear in this region, indicating the probable existence of a solubility limit of La/Th near $x=0.45$. For the Hf, Zr and Ti substitutions, the lattice parameters $a$ and $c$ exhibit behavior similar to Th substitution, wherein the $c$ lattice parameter decreases with increasing  Hf, Zr, and Ti concentration and the $a$ lattice parameter does not show an appreciable concentration dependence. For Sr substitution, the $a$ lattice parameter increases slightly, while $c$ decreases with increasing Sr concentration.

\begin{figure}[b!]
  \includegraphics[width=0.99\linewidth]{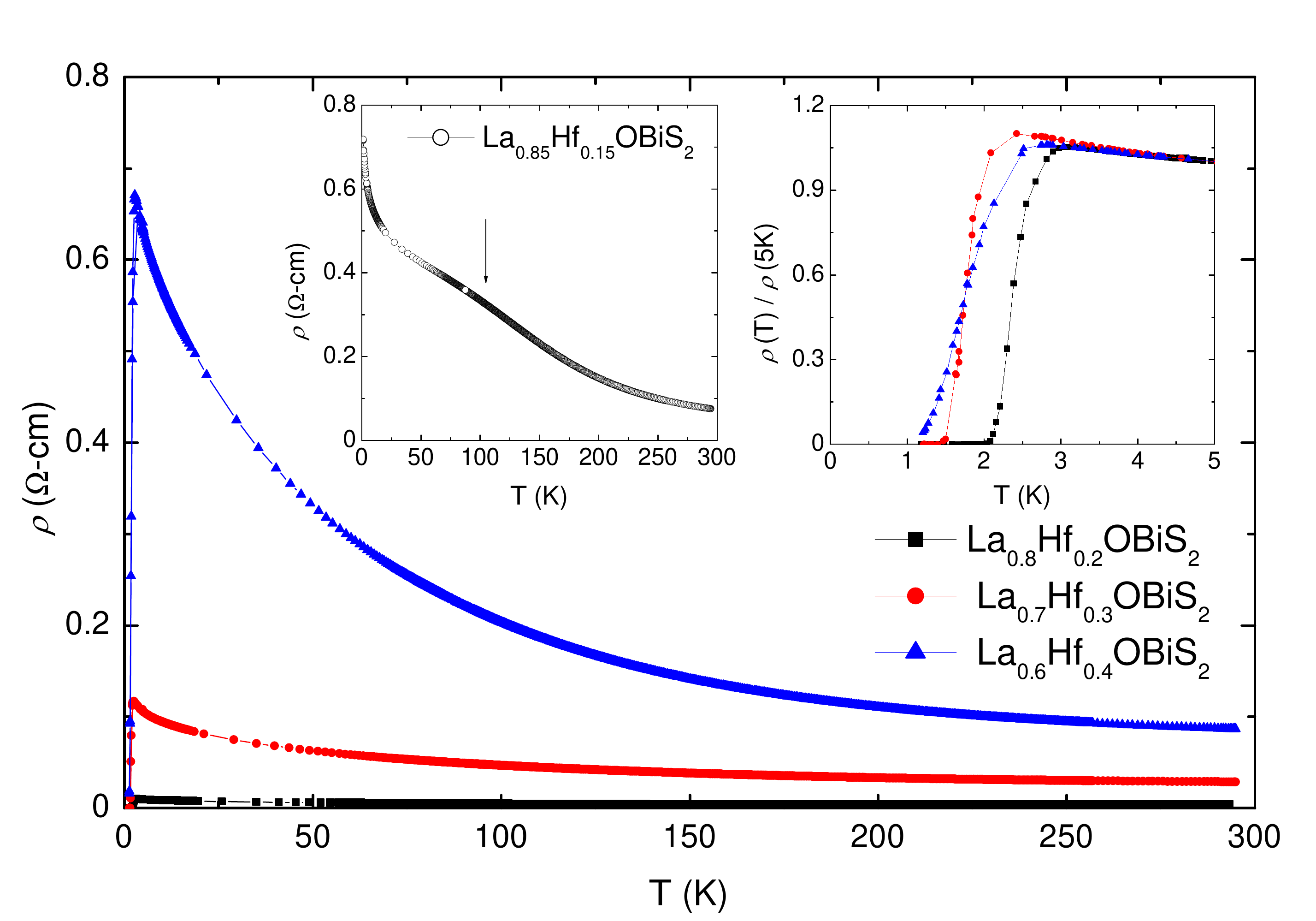}
  \caption{ (Color online) Electrical resistivity $\rho$ vs.~temperature $T$ for \LaHfOBiS. The left inset shows $\rho$ vs.~$T$ for La$_{0.85}$Hf$_{0.15}$OBiS$_2$. The right inset displays the superconducting transition curves for samples with concentrations $0.2 \leq x \leq 0.4$.}\label{fig:rho-Hf}
\end{figure}

\begin{figure}[b!]
  \includegraphics[width=0.99\linewidth]{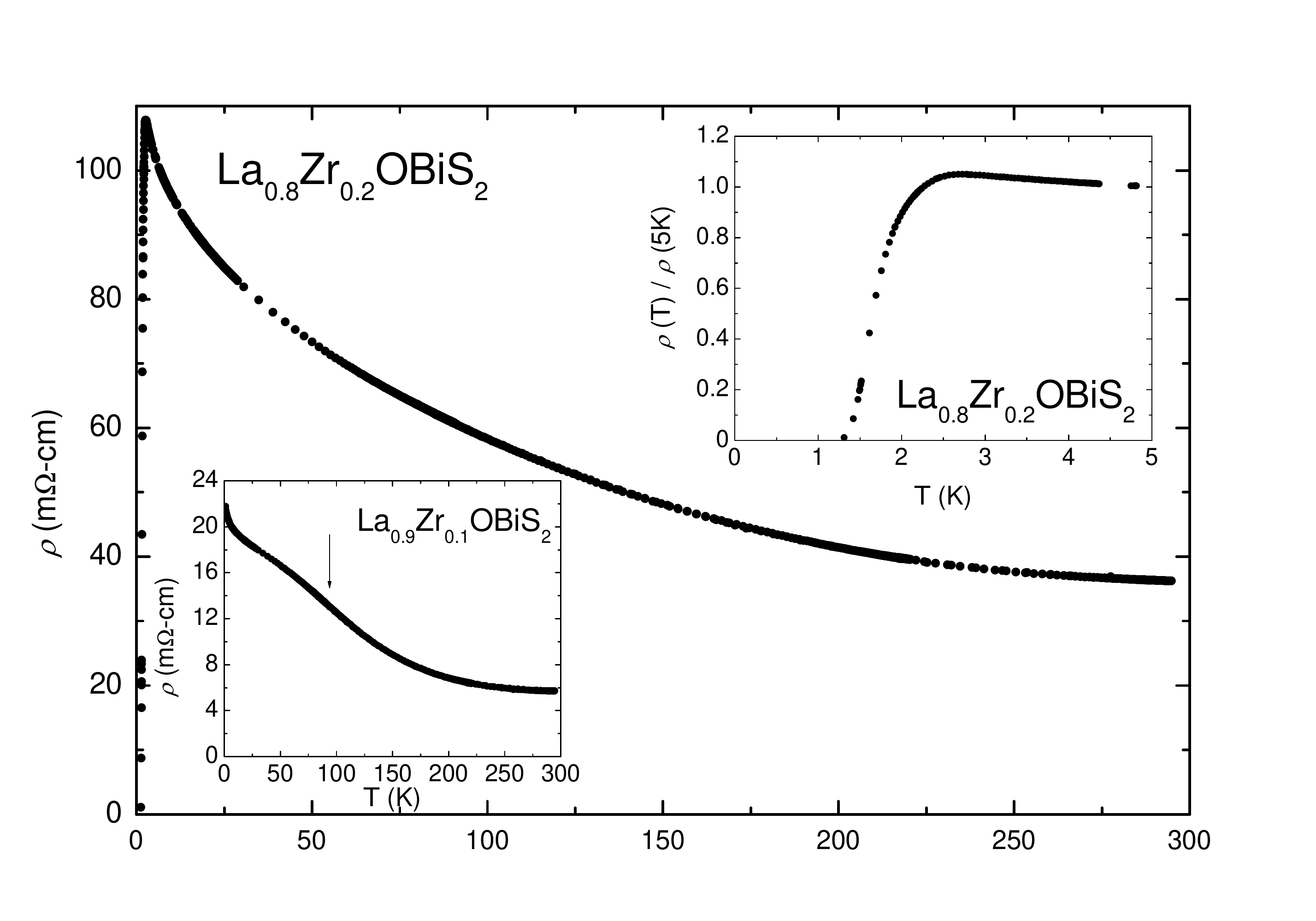}
  \caption{ Electrical resistivity $\rho$ vs.~temperature $T$, plotted for La$_{0.8}$Zr$_{0.2}$OBiS$_2$. The left inset shows $\rho$ vs.~ $T$, for La$_{0.9}$Zr$_{0.1}$OBiS$_2$. The right inset displays the superconducting transition for La$_{0.8}$Zr$_{0.2}$OBiS$_2$.}\label{fig:rho-Zr}
\end{figure}

\begin{figure}[b!]
  \includegraphics[width=0.99\linewidth]{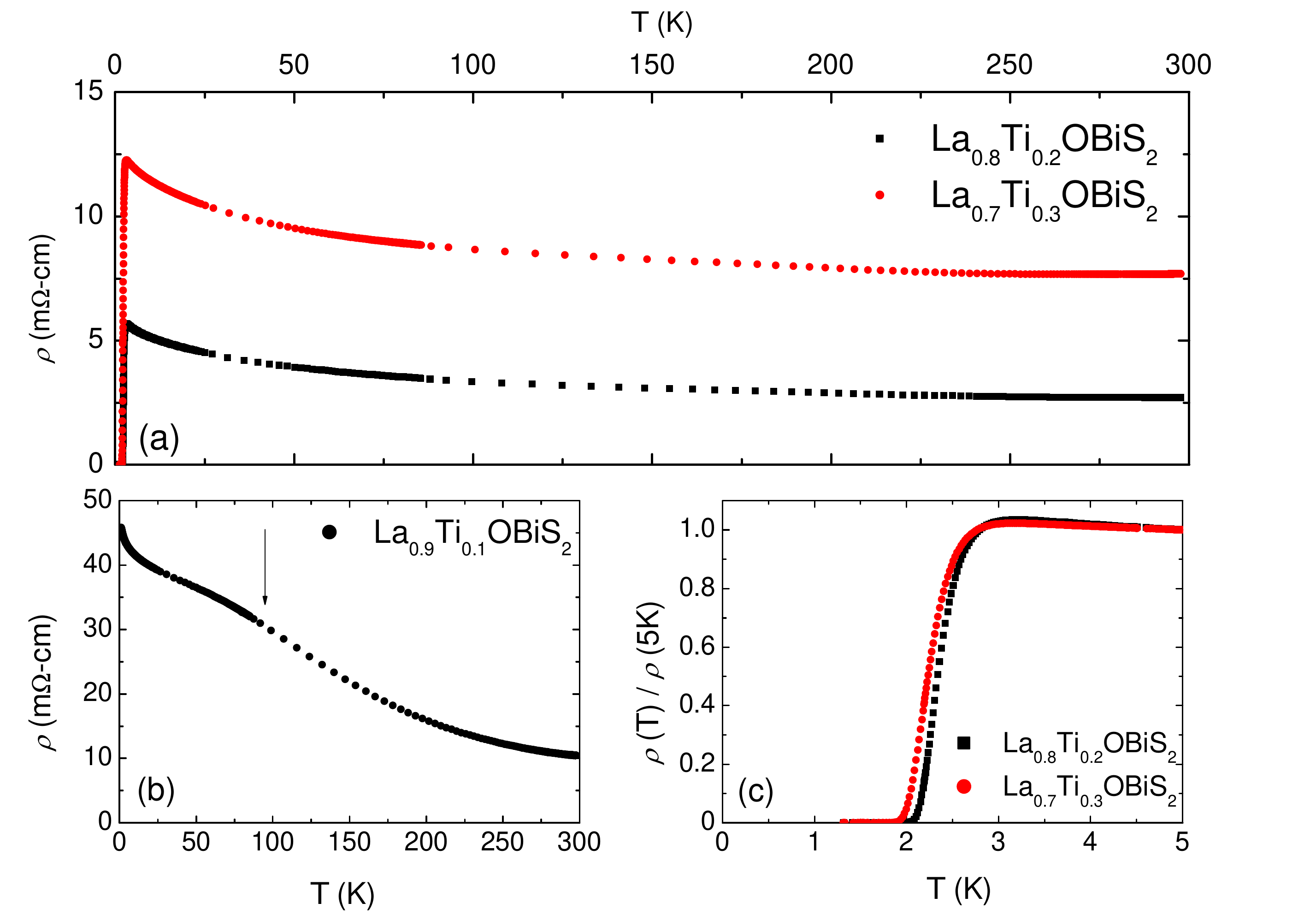}
  \caption{ (Color online) (a) Electrical resistivity $\rho$ vs.~temperature $T$, plotted for \LaTiOBiS~ ($x = 0.2 - 0.3$). (b) $\rho$ vs.~ $T$ for La$_{0.9}$Ti$_{0.1}$OBiS$_2$. (c) The superconducting transition curves for $x = 0.2 - 0.3$.}\label{fig:rho-Ti}
\end{figure}

\begin{figure}[b!]
  \includegraphics[width=0.9\linewidth]{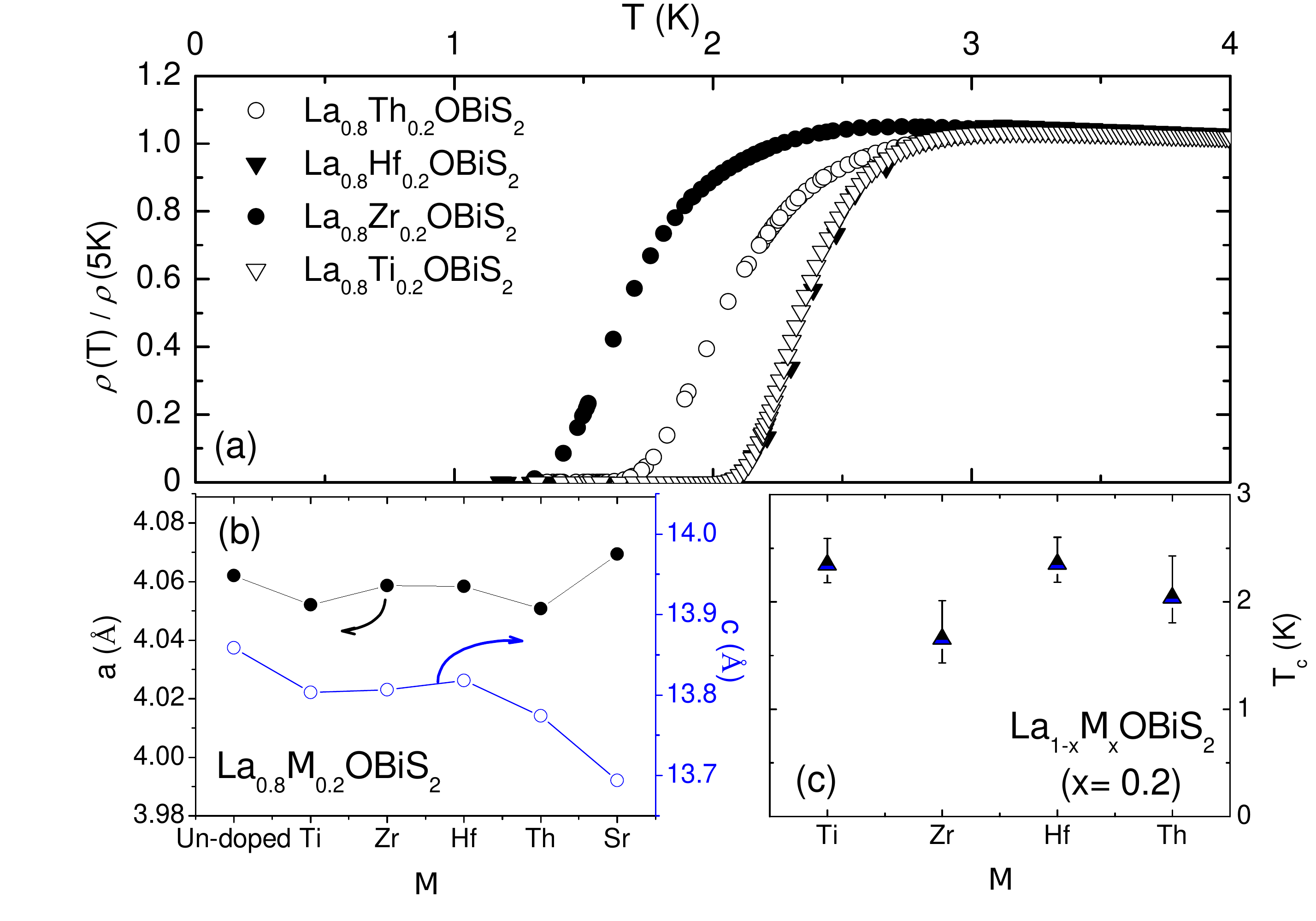}
  \caption{(Color online) (a) The superconducting transition for 20\% concentration of Th, Hf, Zr, and Ti (b) The lattice parameters $a$ and $c$ for doped and un-doped LaOBiS$_2$. (c) Superconducting transition temperatures for La$_{1-x}$$M_x$OBiS$_2$ with \emph{M }= Th, Hf, Zr, Ti and $x = 0.2$.}\label{fig:rho-mix}
\end{figure}

\begin{figure}[b!]
  \includegraphics[width=0.99\linewidth]{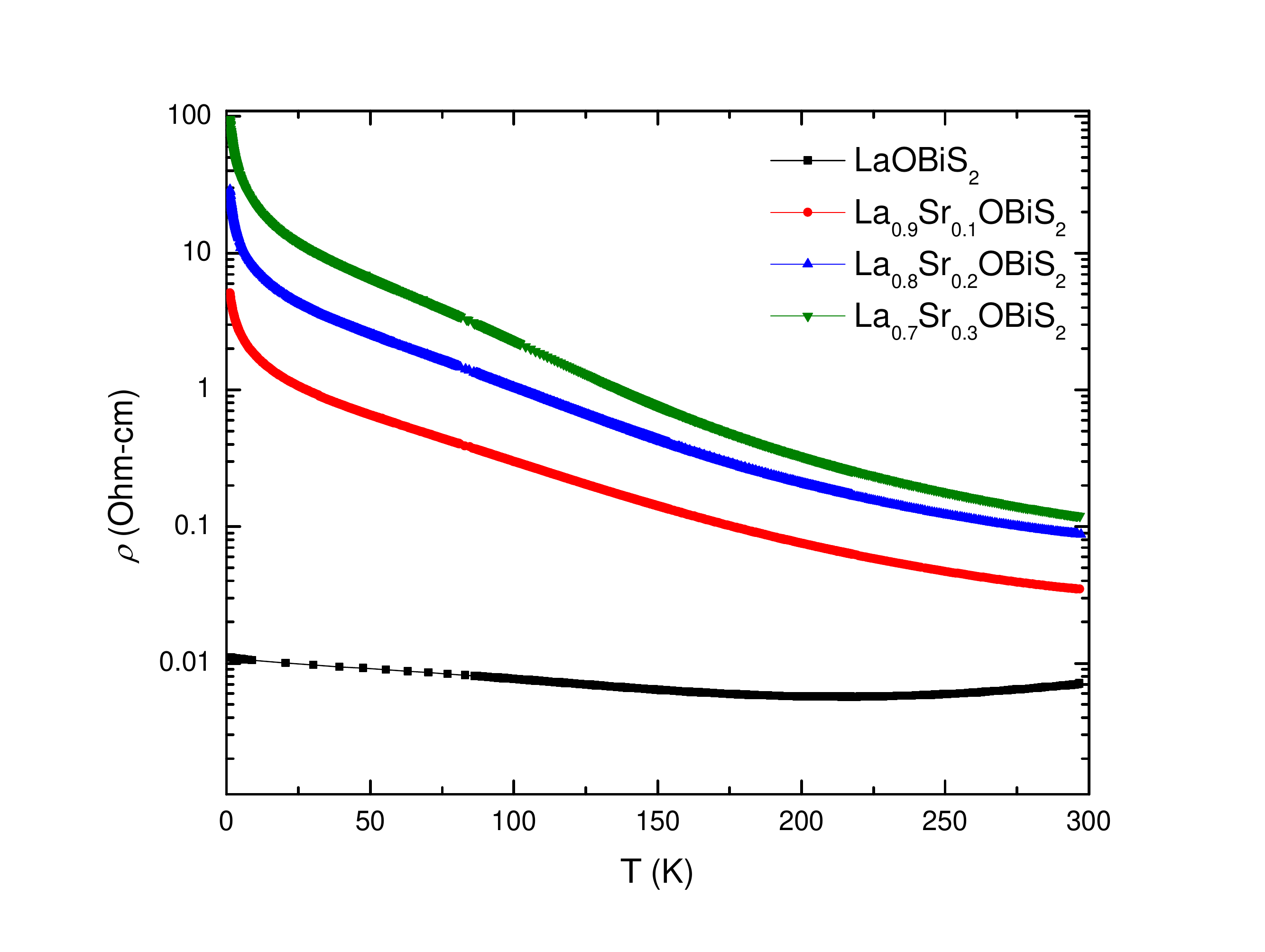}
  \caption{(Color online) Electrical resistivity $\rho$ vs.~temperature $T$, for \LaSrOBiS~ ($ 0 \leq x \leq 0.3$), plotted on a semi-log scale.}\label{fig:rho-Sr}
\end{figure}

\subsection{Electrical resistivity}

Electrical resistivity $\rho(T)$ data, normalized by the value of $\rho$ at 290 K, are shown in Fig.~\ref{fig:rho-Th} for the pure \LaOBiS\ and \ThOBiS\ samples. For both compounds, $\rho$ initially decreases with decreasing temperature, exhibits a minimum at $T$ = 220 K  and $T$ = 206 K, respectively, and then shows semiconductor-like behavior down to the lowest temperatures measured. For the La$_{1-x}$Th$_{x}$OBiS$_2$ system, the minimum in $\rho$ is suppressed for $x = 0.1$, and $\rho(T)$ exhibits relatively strong temperature-dependence and an inflection point (indicated by an arrow) as shown in Fig.~\ref{fig:rho-Th}(b). For $x$ = 0.15, this feature disappears and $\rho$ drops to zero below $T_c$ = 2.85 K. $\rho(T)$ becomes less temperature-dependent for $0.15 \leq x \leq 0.45$ and increases with decreasing temperature until the onset of superconductivity. First principle calculations have suggested that there may be a charge density wave instability or an enhanced correlations in the LaO$_{1-x}$F$_x$BiS$_2$ system. \cite{Yildirim} We are unable to unambiguously observe such instability from the electrical resistivity measurements in this study. However, the inflection point could be related to such an effect.

Electrical resistivity measurements for \LaHfOBiS , ~\LaZrOBiS, and \LaTiOBiS ~samples are shown in Figs.~\ref{fig:rho-Hf}, ~\ref{fig:rho-Zr} and ~\ref{fig:rho-Ti}, respectively. Resistive superconducting transition curves for these systems are indicated in the right inset of Figs.~\ref{fig:rho-Hf}, ~\ref{fig:rho-Zr}, and ~\ref{fig:rho-Ti}(c). All of these doping studies show similar characteristics such as an inflection point anomaly in the normal state with low concentration and induced superconductivity for concentrations starting with $x$ = 0.15 for \LaThOBiS ~and ~$x$ = 0.2 for \LaHfOBiS , \LaZrOBiS, and \LaTiOBiS. Furthermore, all of these systems show semiconductor-like behavior in the normal state. The inflection point in $\rho$ of samples with low concentration is emphasized in the left inset of Figs.~\ref{fig:rho-Hf}, ~\ref{fig:rho-Zr}, and in Fig.~\ref{fig:rho-Ti}(b) where the anomaly is indicated by an arrow. These anomalies all appear to be present at a common temperature of roughly 120 K in concentrations $x$ just below those where superconductivity is induced. To compare the superconducting transition temperatures ($T_c$) we consider data for $20\%$ substitution of La by Th, Hf, Zr, and Ti in Fig.~\ref{fig:rho-mix}(a). The superconducting transition temperatures, as shown in Fig.~\ref{fig:rho-mix}(c), are characterized by the temperatures where the electrical resistivity drops to $50\%$ of the normal state resistivity, and the width of the transition is determined by the temperatures where the resistivity drops to $90\%$ and $10\%$ of the normal state resistivity. Electron doping clearly induces superconductivity in LaOBiS$_2$. The $T_c$'s are quite similar to one another, but the transition width is sharper for La$_{0.8}$Hf$_{0.2}$OBiS$_2$ and La$_{0.8}$Ti$_{0.2}$OBiS$_2$ than for La$_{0.8}$Zr$_{0.2}$OBiS$_2$ and La$_{0.8}$Th$_{0.2}$OBiS$_2$. The lowest $T_c$ is seen in La$_{0.8}$Zr$_{0.2}$OBiS$_2$. There does not appear to be a clear correlation between $T_c$ and the lattice parameters, shown in Fig.~\ref{fig:rho-mix}(b). Meanwhile,  Fig.~\ref{fig:rho-Sr} shows $\rho(T)$ measurements for \LaSrOBiS\, wherein no evidence of a superconducting transition is observed down to $\sim$ 1 K in the range $ 0.1 \leq x \leq 0.3$. This result suggests that hole doping is not sufficient to induce superconductivity. It is, however, interesting to note that the magnitude of $\rho$ at low temperatures increases with increasing Sr concentration, which is similar to the behavior observed with Th, Hf, Zr, and Ti doping.

\begin{figure}[th!]
  \includegraphics[width=0.99\linewidth]{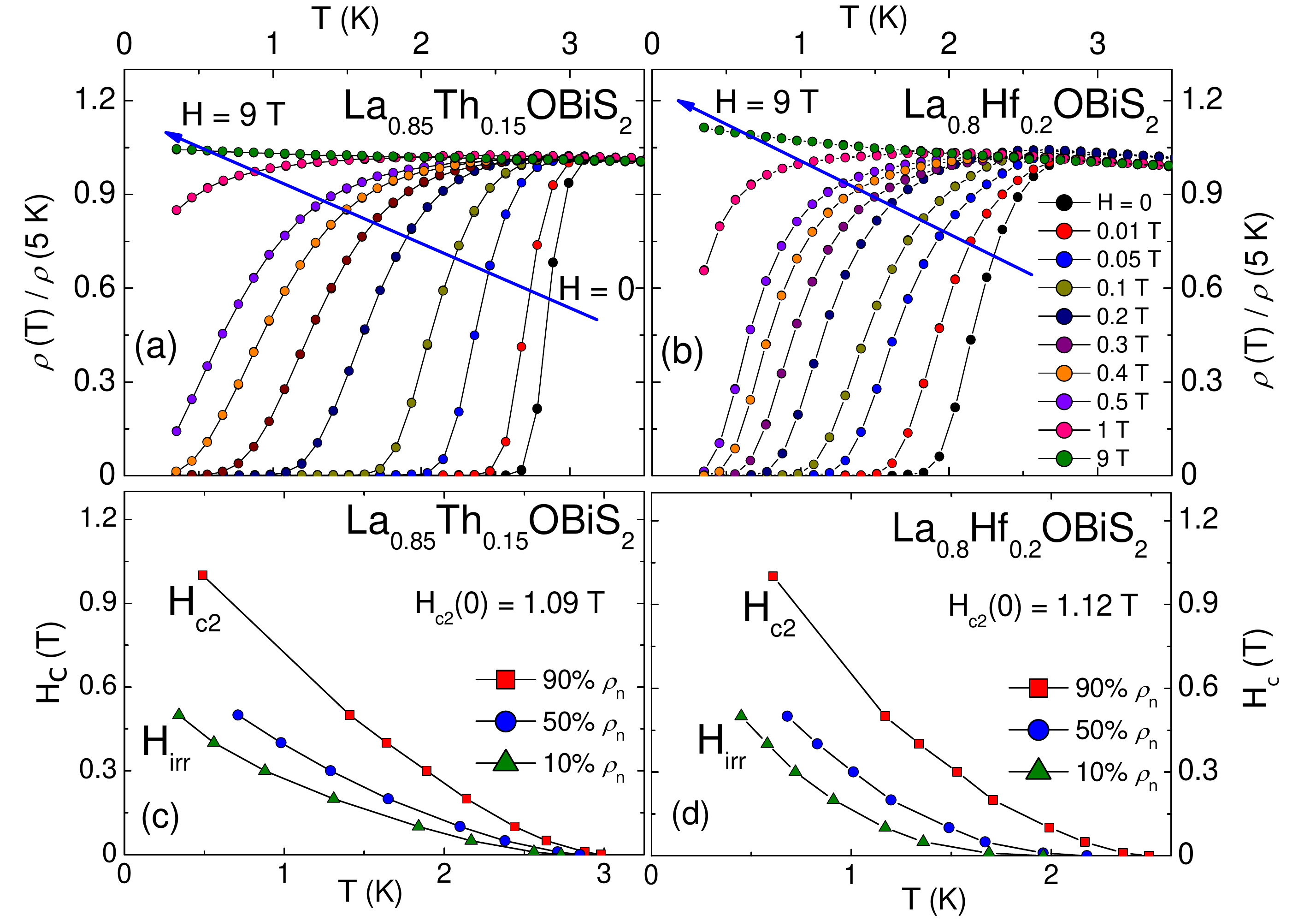}
  \caption{(Color online) (a, b) Resistive superconducting transition curves for La$_{0.85}$Th$_{0.15}$OBiS$_2$ and La$_{0.8}$Hf$_{0.2}$OBiS$_2$ samples, respectively, measured under several different applied magnetic fields ($H =$ 0, 0.01, 0.05, 0.1, 0.2, 0.3, 0.4, 0.5, 1, 9 T). (c, d) The temperature dependence of the upper critical field $H_{c2}$, and $H_{irr}$, determined from the $90\%$ and $10\%$ normal state $\rho$ for La$_{0.85}$Th$_{0.15}$OBiS$_2$ and La$_{0.8}$Hf$_{0.2}$OBiS$_2$ samples, respectively. The temperature corresponding to the $50\%$ normal state $\rho$ is also shown.}\label{fig:magnetoresistance}
\end{figure}

The temperature dependence of $\rho$, normalized by its value at 5 K for La$_{0.85}$Th$_{0.15}$OBiS$_2$ and La$_{0.8}$Hf$_{0.2}$OBiS$_2$ samples, is shown in Fig.~\ref{fig:magnetoresistance}(a, b), respectively, under several applied magnetic fields ($H =$ 0, 0.01, 0.05, 0.1, 0.2, 0.3, 0.4, 0.5, 1, and 9 T) down to 0.36 K. Both samples undergo relatively sharp superconducting transitions at $T_c$ = 2.85 K and  $T_c$ = 2.40 K for La$_{0.8}$Th$_{0.2}$OBiS$_2$ and La$_{0.8}$Hf$_{0.2}$OBiS$_2$, respectively. With increasing magnetic field, the transition width broadens and the onset of superconductivity gradually shifts to lower temperatures. Similar broadening of the transition was observed in the high-$T_c$ layered cuprate and Fe-pnictide superconductors and attributed to the vortex-liquid state.\cite{Safar, Lee}  Fig.~\ref{fig:magnetoresistance}(c, d) shows the upper critical field $H_{c}$ vs.~$T$ for La$_{0.85}$Th$_{0.15}$OBiS$_2$ and La$_{0.8}$Hf$_{0.2}$OBiS$_2$ samples, corresponding to the temperatures where the resistivity drops to $90\%$ of the normal state resistivity $\rho_n(T,H)$ ($T_{c, onset}$), $50\%$ of $\rho_n(T,H)$ ($T_c$), and $10\%$ of $\rho_n(T,H)$ ($T_{c, zero}$) in applied magnetic fields. Using the conventional one-band Werthamer-Helfand-Hohenberg (WHH) theory, \cite{Werthamer} the orbital critical fields $H_{c2}(0)$ for La$_{0.85}$Th$_{0.15}$OBiS$_2$ and La$_{0.8}$Hf$_{0.2}$OBiS$_2$ compounds were inferred from their initial slopes of $H_{c}$ with respect to \emph{T}, yielding values of 1.09 T and 1.12 T, respectively.  These values of $H_{c2}(0)$, are very close to the values seen in Sr$_{1-x}$La$_x$FBiS$_2$ (1.04 T) \cite{Lin} and LaO$_{0.5}$F$_{0.5}$BiS$_2$ (1.9 ~T), \cite{Awana} suggesting that the superconducting state in BiS$_2$-based superconductors probably shares a common character.

\begin{figure}[th!]
  \includegraphics[width=0.99\linewidth]{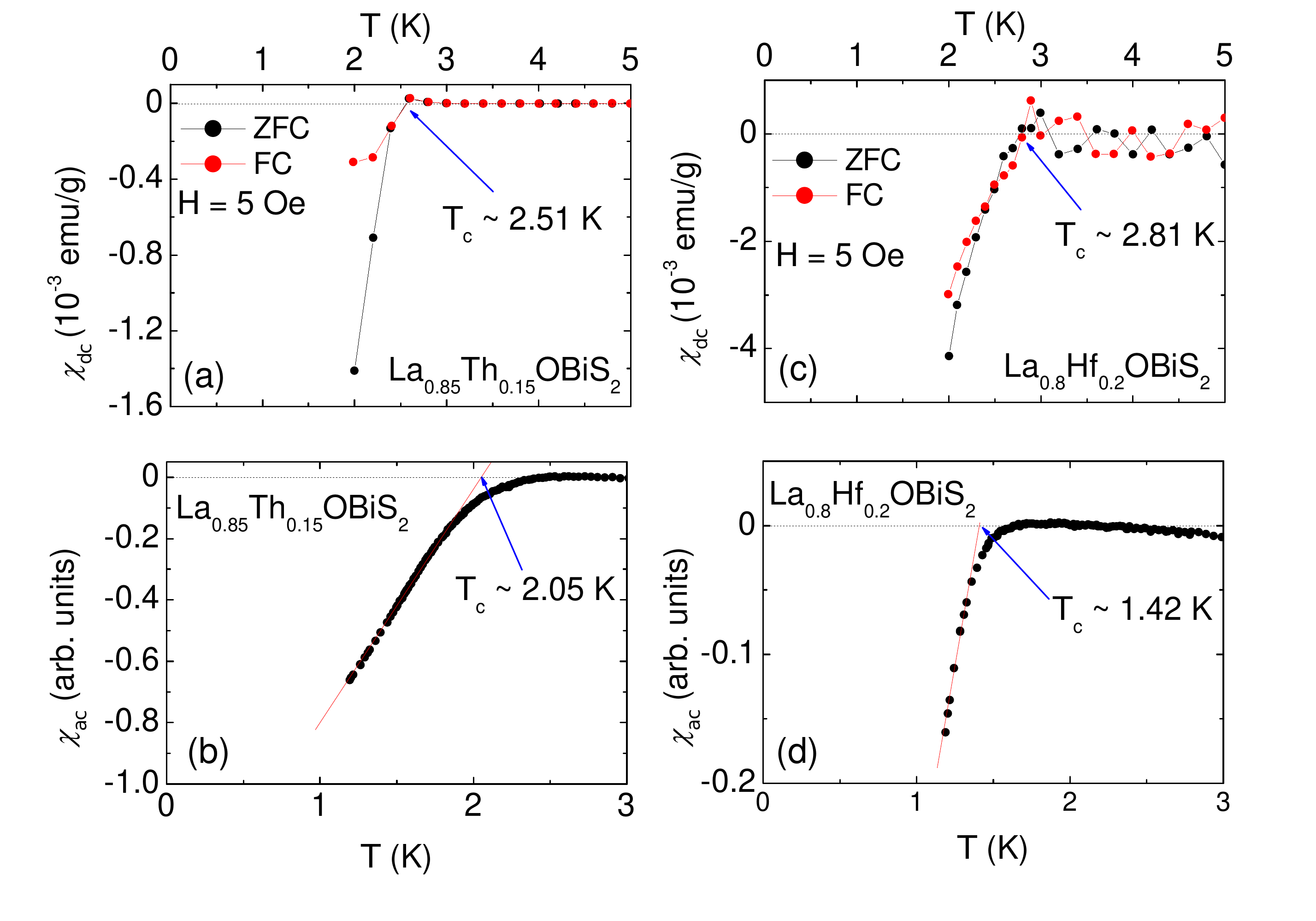}
  \caption{(Color online) (a, c) DC magnetic susceptibility $\chi_{dc}$ vs. temperature $T$ for La$_{0.85}$Th$_{0.15}$OBiS$_2$ and La$_{0.8}$Hf$_{0.2}$OBiS$_2$, respectively, measured in field cooled (FC) and zero field cooled (ZFC) conditions with a 5 Oe applied magnetic field. (b, d) AC magnetic susceptibility $\chi_{ac}$ vs. $T$ for La$_{0.85}$Th$_{0.15}$OBiS$_2$ and La$_{0.8}$Hf$_{0.2}$OBiS$_2$, respectively. The superconducting critical temperature $T_c$ is indicated and labeled explicitly. }\label{fig:M}
\end{figure}

\begin{figure}[t!]
  \includegraphics[width=0.99\linewidth]{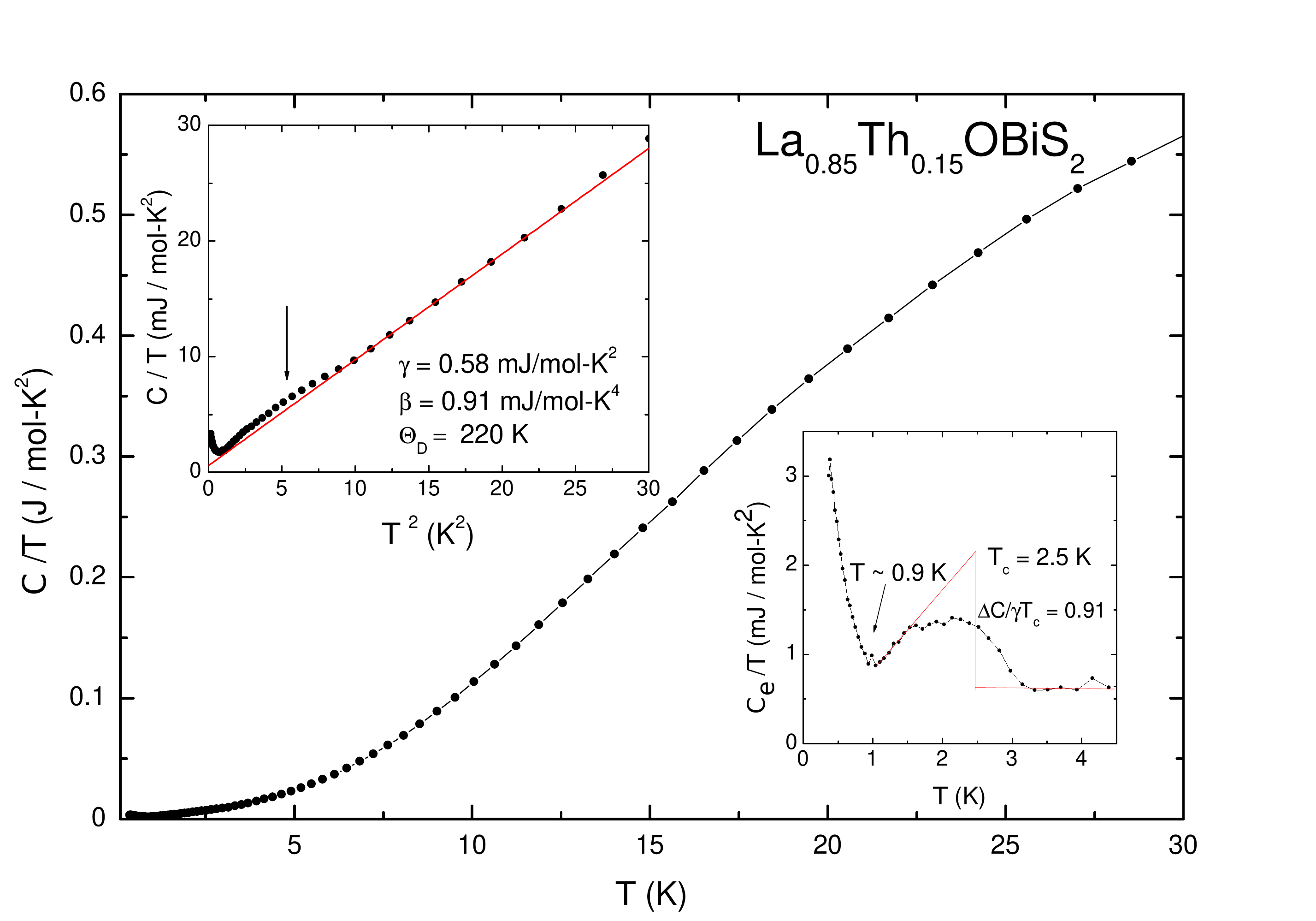}
  \caption{Specific heat $C$ divided by temperature $T$, $C / T$  vs. temperature $T$ for La$_{0.85}$Th$_{0.15}$OBiS$_2$.  $C / T$ vs. $T^2$ is shown in the inset in the upper left hand part of the figure.  The red line is a fit of the expression $C(T) = \gamma T + \beta T^{3}$ to the data which yields  $\gamma = 0.58$  mJ/mol-K$^2$ and  $\Theta_D$ = 220 K. The inset in the lower right part of the figure shows a plot of $C_e$ vs. $T$, where $C_e / T$ is the electronic contribution to the specific heat, in the vicinity of the superconducting transition. An idealized entropy conserving construction yields $T_c$ = 2.5 K and  $\Delta C / \gamma T_c$ = 0.91.}\label{fig:C-Th}
\end{figure}

\begin{figure}[t!]
  \includegraphics[width=0.99\linewidth]{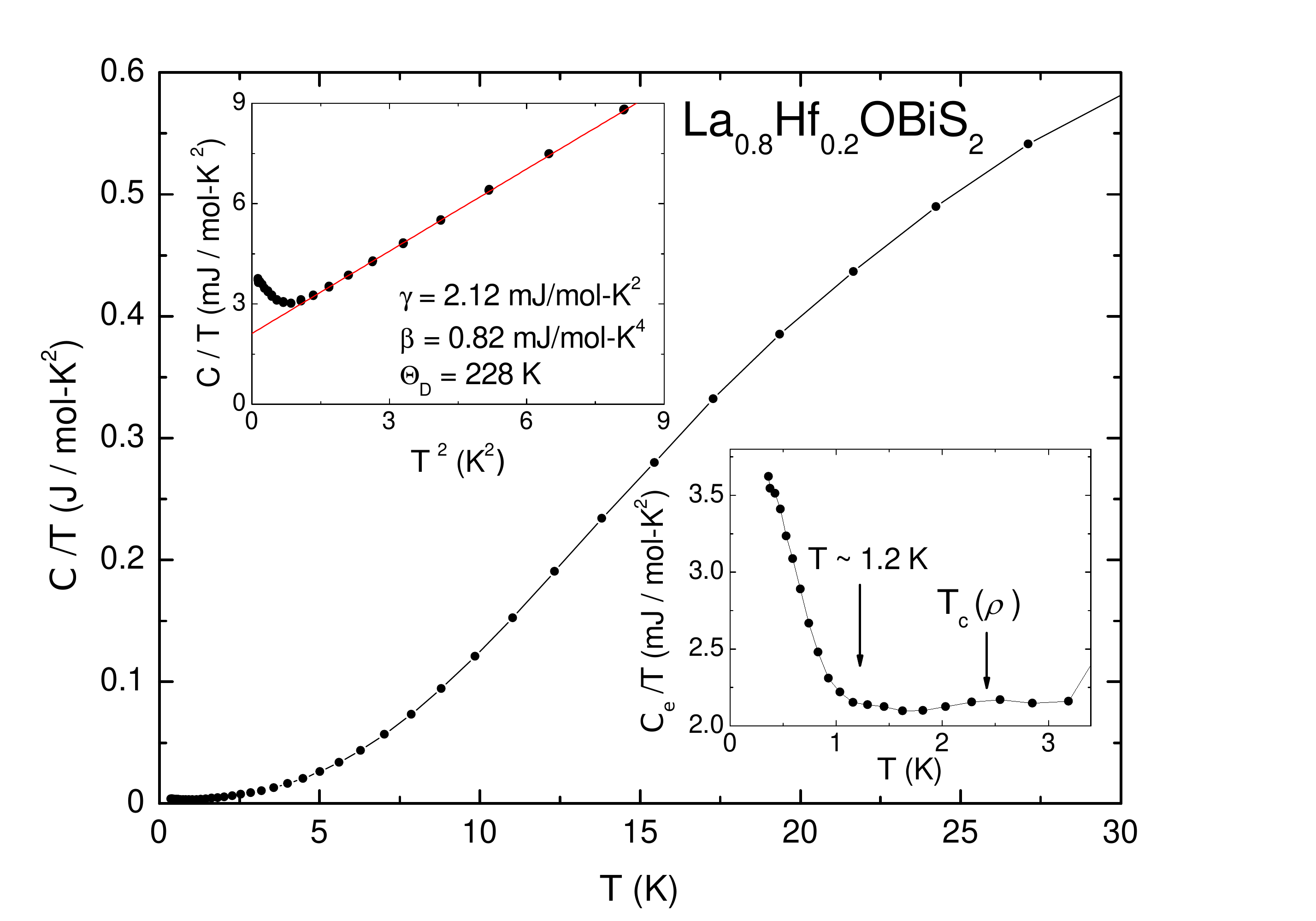}
  \caption{Specific heat $C$ divided by temperature $T$, $C / T$  vs. temperature $T$  La$_{0.8}$Hf$_{0.2}$OBiS$_2$. A plot of $C / T$ vs. $T^2$ is shown in the inset in the upper left hand part of the figure.  The red line is a fit of the expression $C(T) = \gamma T + \beta T^{3}$ to the data which yields $\gamma = 2.12$  mJ/mol-K$^2$ and  $\Theta_D$ = 228 K. A plot of $C_e / T$ vs. $T$, in the vicinity of the superconducting transition in the inset in the lower right hand side of the figure. The superconducting transition temperature $T_c$, obtained from the electrical resistivity measurements ($T_c$ = 2.36 K), is indicated by an arrow.}\label{fig:C-Hf}
\end{figure}

\subsection{Magnetization}

To determine whether superconductivity is a bulk phenomenon in La$_{1-x}M_x$OBiS$_2$, zero field cooled (ZFC) and field cooled (FC) measurements of the magnetic susceptibility  $\chi_{dc}(T)$  were made in a magnetic field of 5 Oe for the La$_{0.85}$Th$_{0.15}$OBiS$_2$ and La$_{0.8}$Hf$_{0.2}$OBiS$_2$ samples. These measurements are plotted in Fig.~\ref{fig:M}(a, c). ZFC measurements yield diamagnetic screening signals with $T_c$ onset values that are lower than values obtained from $\rho(T)$ data, while FC measurements reveal evidence for strong vortex pinning. AC magnetic susceptibility measurements for La$_{0.85}$Th$_{0.15}$OBiS$_2$ and La$_{0.8}$Hf$_{0.2}$OBiS$_2$ samples are shown in Fig.~\ref{fig:M}(b, d), respectively. These data exhibit signatures of SC with slightly lower $T_c$'s than those observed in $\rho$ and $\chi_{dc}$ measurements. It is clear that, even though the superconducting transition is incomplete at 1.3 K, the volume fraction at that temperature is appreciable for both samples.

\subsection{Specific heat}

Specific heat data for La$_{0.85}$Th$_{0.15}$OBiS$_2$ and La$_{0.8}$Hf$_{0.2}$OBiS$_2$ samples are displayed in Fig.~\ref{fig:C-Th} and Fig.~\ref{fig:C-Hf}, respectively. The data for La$_{0.85}$Th$_{0.15}$OBiS$_2$ cover the temperature range 0.36 K to 30 K.  A fit of the expression $C(T) = \gamma T + \beta T^{3}$ to the data in the normal state, where  $\gamma$ is the electronic specific heat coefficient and  $\beta$ is the coefficient of the lattice contribution, is plotted as a function of  $T^2$ in the inset in the upper left hand side of Fig.~\ref{fig:C-Th}. From the best fit, which is explicitly indicated by a line in the inset, we obtain values of $\gamma = 0.58$  mJ/mol-K$^2$ and  $\beta = 0.91$ mJ/mol-K$^4$; the value of $\beta$ corresponds to a Debye temperature of  $\Theta_D = 220$ K. In the inset in the lower right hand side of Fig.~\ref{fig:C-Th}, $C_e / T$ vs. ~$T$ is plotted. A clear feature is observed between $1$ and $3$ K. If this feature is related to the transition into the superconducting state, we can estimate $T_c$ = 2.5 K from an idealized entropy conserving construction. This value of $T_c$ is close to the temperature obtained from the electrical resistivity ($T_c$ = 2.85 K). The presence of the feature may suggest that superconductivity is a bulk phenomenon in this compound.  The ratio of the jump to $\gamma$ $T_c$, $\Delta C / \gamma T_c$ = 0.91, was calculated using a jump in $C_e / T$ of 0.53 mJ/mol-K$^2$, extracted from the entropy conserving construction as seen in the inset in the lower right hand side of the figure. This value  is less than the value of 1.43 predicted by the BCS theory, but is similar to that seen in LaO$_{0.5}$F$_{0.5}$BiS$_2$.\cite{Yazici}

The specific heat data for La$_{0.8}$Hf$_{0.2}$OBiS$_2$ are displayed between 0.36 K and 30 K in Fig.~\ref{fig:C-Hf}. The upper inset of Fig.~\ref{fig:C-Hf} highlights the linear fit to the $C/T$ data plotted vs. $T^2$, from which we extracted $\gamma = 2.12$  mJ/mol-K$^2$ and $\beta =  0.82$ mJ/mol-K$^4$, and calculated  $\Theta_D = 228$ K. In the inset in the lower right hand side of the figure, the electronic contribution to the specific heat $C_e / T$ vs. ~$T$ is shown, which has been obtained by subtracting the lattice contribution $\beta T^{3}$ from $C(T)$. Absent from these data is any clear evidence for a jump at the $T_c$ obtained from either the electrical resistivity or DC and AC magnetic susceptibility ($T_c$ = 2.36 K, 2.81 K and 1.42 respectively) measurements.  However, there is a small feature around the $T_c$ obtained from the electrical resistivity ($T_c$ = 2.36 K) as indicated by an arrow in the lower right hand side of the figure. The absence of a well-defined superconducting jump at  $T_c$ is probably a consequence of the superconducting transition being spread out in temperature due to sample inhomogeneity.

There is an upturn in specific heat below roughly 1.2 K and 0.9 K for the La$_{0.8}$Hf$_{0.2}$OBiS$_2$ and La$_{0.85}$Th$_{0.15}$OBiS$_2$ samples, respectively. The same kind of upturn is also seen in Sr$_{0.5}$La$_{0.5}$FBiS$_2$ at a similar temperature. This upturn may be a contribution to specific heat from a Schottky anomaly or may be indicative of a second superconducting phase in this compound.\cite{Lin} However, measurements must be made at temperatures below 0.36 K to unambiguously clarify the nature of this feature.

\section{Discussion}\label{sec:discussion}

\begin{figure}[t]
  \includegraphics[width=0.99\linewidth]{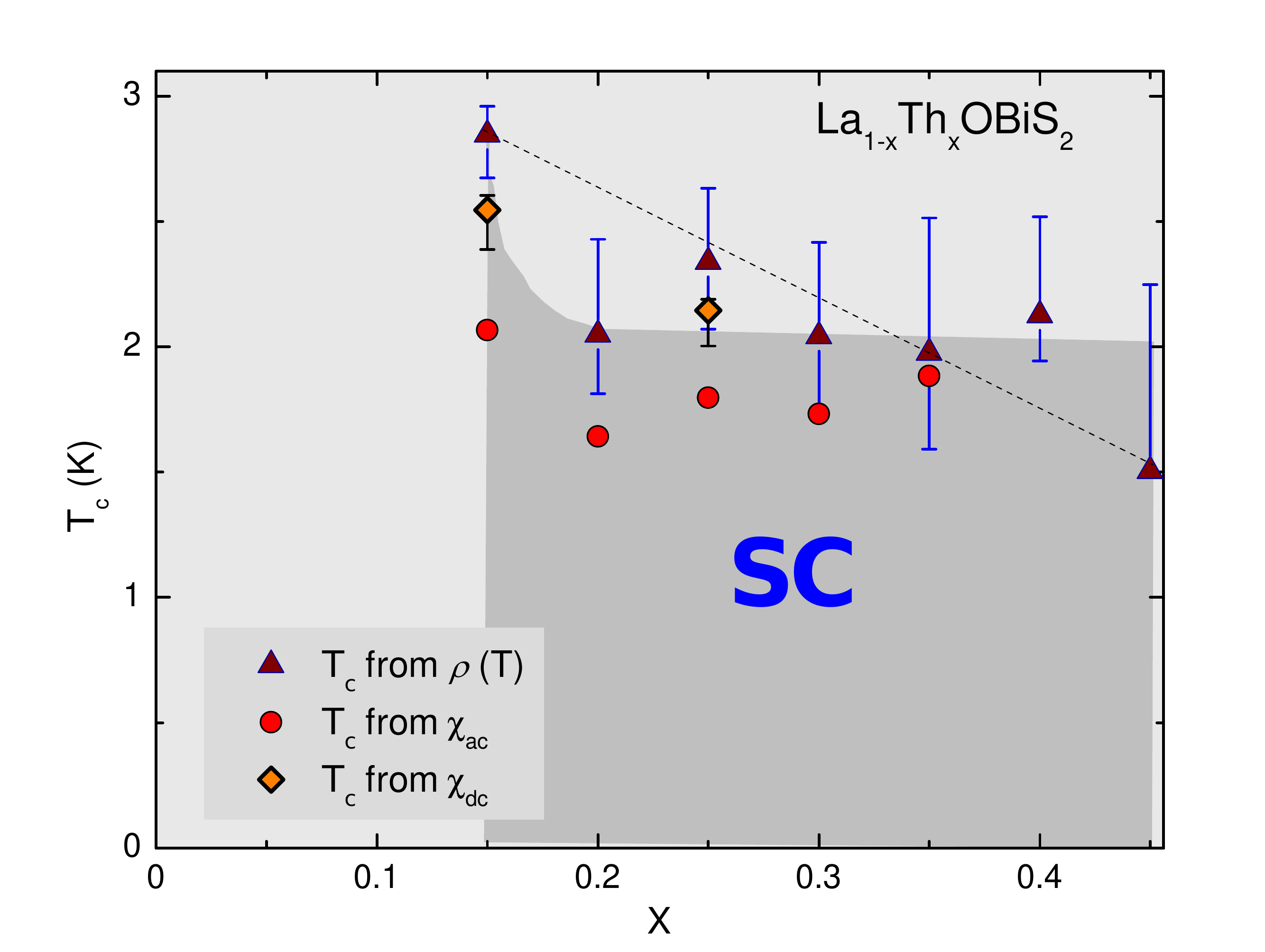}
  \caption{(Color online) (a) Superconducting transition temperature $T_c$ vs. Th concentration $x$ phase diagram for the \LaThOBiS ~system. The phase diagram is constructed from electrical resistivity $\rho$ (filled triangles), DC magnetic susceptibility $\chi_{dc}$ (filled diamonds), and AC magnetic susceptibility $\chi_{ac}$ (filled circles) measurements. The $T_c - x$ phase boundary (dark gray area) separates the superconducting phase (SC) from the normal phase. The value of $T_c$ from $\rho (T)$ was defined by the temperature where $\rho$ drops to $50\%$ of its normal state value and vertical bars indicate the temperatures where the electrical resistivity drops to $90\%$ and $10\%$ of its normal state value. Values of $T_c$ from $\chi_{ac}$ and  $\chi_{dc}$ were defined by the onset of the superconducting transition. The dashed line indicates the another possible upper bound for the SC region}\label{fig:phasediagram}
\end{figure}

The results from $\rho(T)$, $\chi_{dc}$, and $\chi_{ac}$ measurements are summarized in a phase diagram of $T_c$ vs. Th concentration $x$ shown in Fig.~\ref{fig:phasediagram}. $T_c(x)$ decreases with $x$ from 2.85 K at $x$ = 0.15 to 2.05 K at $x$ = 0.20 and exhibits roughly concentration-independent behavior at higher concentration. The superconducting region may be defined by the dark gray region in Fig.~\ref{fig:phasediagram}, and apparently lacks a dome-like character typically seen for both the high-$T_c$ layered cuprate and Fe-pnictide superconductors. Dome-like superconductor regions are also seen in LaO$_{1-x}$F$_x$BiS$_2$, and NdO$_{1-x}$F$_x$BiS$_2$. \cite{Deguchi, Demura} At concentrations below the SC domes in these compounds, electrical resistivity measurements reveal bad metal or semiconducting-like behavior. These results are in agreement with first principle calculations \cite{Yildirim} which suggest that the density of states at the Fermi level increases with increasing electron doping, such that the insulating parent compound becomes metallic. This effect is expected to be maximal at half filling ($x$ = 0.5). The LaO$_{1-x}$F$_x$BiS$_2$ system \cite{Deguchi} shows a maximum $T_c$ for $x$ = 0.5 while the NdO$_{1-x}$F$_x$BiS$_2$ \cite{Demura} system exhibits its highest $T_c$ at $x$ = 0.3. Other scenarios may be possible depending on how we define the SC region because of the broadness of the superconducting transitions. For example, a dashed line is also shown in Fig.~\ref{fig:phasediagram} which mostly resides within the ranges characterized by the transition. In order to better define the phase boundary, studies on samples with sharper transitions would be beneficial.

%$T_c$ is dependent on the sample preperation technique in BiS$_2$-based superconductors. (LaO$_{0.5}$F$_{0.5}$BiS$_2$ compound $T_c$ reaches up to $10$ ~K when it prepared under pressure, \cite{Mizuguchi2, Deguchi} when it prepared at ambient pressure $T_c$ is much more lower then $10$ ~K. \cite{Mizuguchi2, Deguchi, Awana} However, the mechanism of this procedure has not well understood. It is may be related with internal strain, may be the size effect or may be it influences the oxygen content. For these reasons, this unusual phase diagram Fig.~\ref{fig:phasediagram} may be related to the non-optimized samples.

\begin{figure}[t]
  \includegraphics[width=0.99\linewidth]{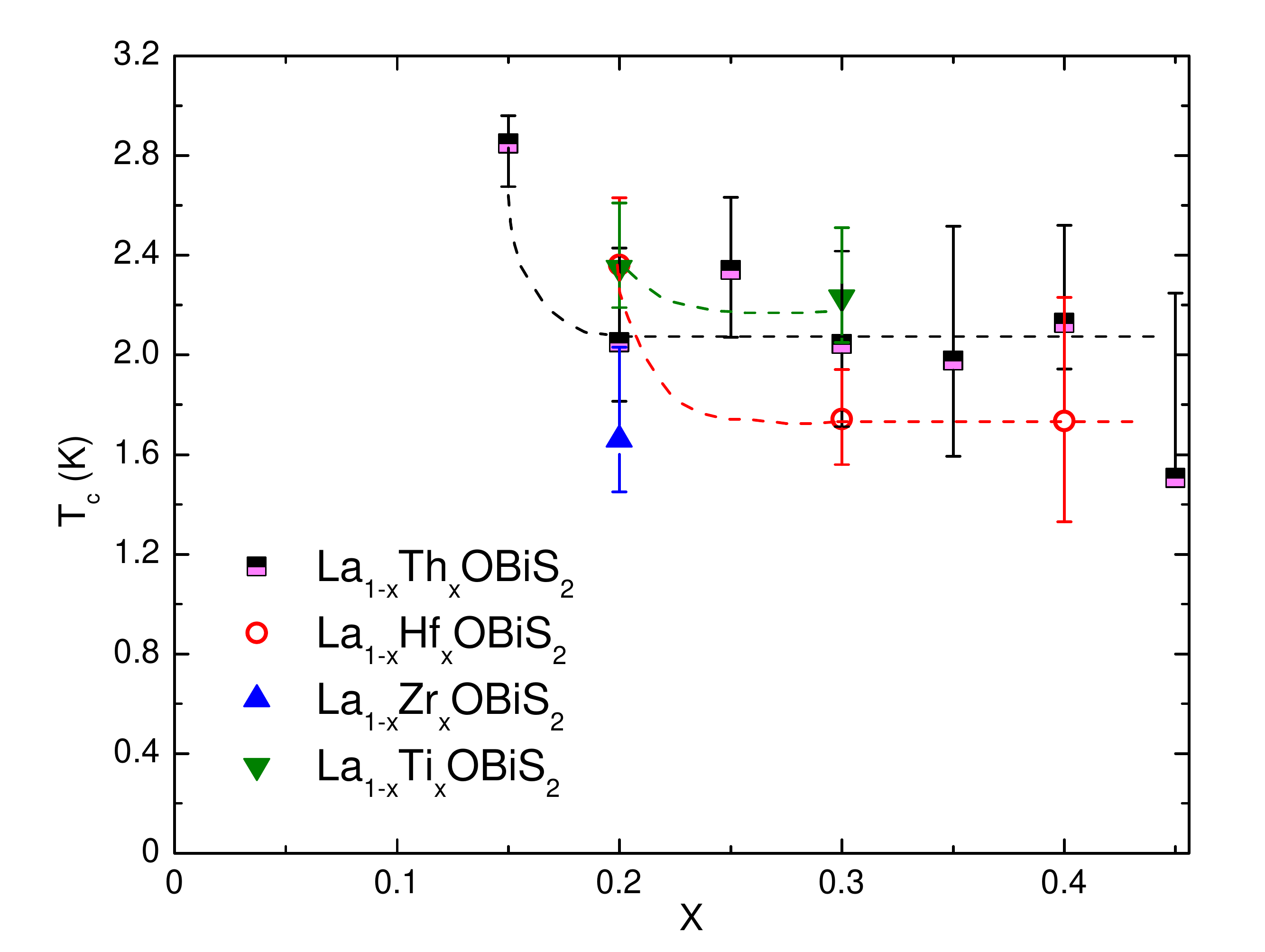}
  \caption{(Color online) Superconducting transition temperature $T_c$ vs. concentration $x$ phase diagram for the La$_{1-x}$$M_{x}$OBiS$_2$ ~system ~(\emph{M} = Th, Hf, Zr, Ti), constructed from electrical resistivity ($\rho (T)$) measurements. The value of $T_c$ was defined by the temperatures where  $\rho$ drops to $50\%$ of its normal state value and vertical bars are defined by the temperatures where the electrical resistivity drops to $90\%$ and of $10\%$ of its normal state value.}\label{fig:phasediagram-mix}
\end{figure}

A similar result can be seen in Fig.~\ref{fig:phasediagram-mix} where the highest $T_c(x)$ is observed at $x$ = 0.2 for Hf, Zr and Ti doping. $T_c(x)$ decreases initially with increasing $x$ and then becomes concentration-independent. Since the ionic sizes of Th, Hf, Zr, and Ti are similar, the $T_c$' s are also very close to each other. The character of $T_c(x)$ observed for La$_{1-x}M_x$OBiS$_2$ and displayed in the phase diagram suggests that the superconducting state is similar for these systems.  Considering other examples of inducing superconductivity by electron doping such as in LaO$_{1-x}$F$_x$BiS$_2$\cite{Yazici, Mizuguchi2, Jha, Deguchi, Awana, Demura} and Sr$_{1-x}$La$_x$FBiS$_2$,\cite{Lin} our results suggest that electron doping is a viable approach to induce superconductivity in BiS$_2$ based compounds.

\section{Concluding remarks}

In summary, we have synthesized polycrystalline samples of La$_{1-x}$$M_{x}$OBiS$_2$ ($M$ = Th, Hf, Zr, Ti, and Sr) with the ZrCuSiAs crystal structure. Electrical resistivity, DC and AC magnetic susceptibility, and specific heat measurements were performed on selected samples. Electron doping via substitution of tetravalent Th$^{+4}$, Hf$^{+4}$, Zr$^{+4}$, Ti$^{+4}$ for trivalent La$^{+3}$ induces superconductivity while hole doping via substitution of divalent Sr$^{+2}$ for La$^{+3}$ does not. These results strongly suggest that electron doping by almost any means may be sufficient to induce superconductivity in BiS$_2$-based compounds.

\begin{acknowledgments}
The authors gratefully acknowledge the support of the US Air Force Office of Scientific Research - Multidisciplinary University Research Initiative under Grant No. FA9550-09-1-0603 and the US Department of Energy under Grant No. DE-FG02-04-ER46105.
\end{acknowledgments}

%\bibliographystyle{prsty}
%\begin{thebibliography}{10}

\bibliography{La1-xThxOBiS2}

\begin{thebibliography}{40}
\expandafter\ifx\csname natexlab\endcsname\relax\def\natexlab#1{#1}\fi
\expandafter\ifx\csname bibnamefont\endcsname\relax
  \def\bibnamefont#1{#1}\fi
\expandafter\ifx\csname bibfnamefont\endcsname\relax
  \def\bibfnamefont#1{#1}\fi
\expandafter\ifx\csname citenamefont\endcsname\relax
  \def\citenamefont#1{#1}\fi
\expandafter\ifx\csname url\endcsname\relax
  \def\url#1{\texttt{#1}}\fi
\expandafter\ifx\csname urlprefix\endcsname\relax\def\urlprefix{URL }\fi
\providecommand{\bibinfo}[2]{#2}
\providecommand{\eprint}[2][]{\url{#2}}

\bibitem[{\citenamefont{Mizuguchi
  et~al.}(2012{\natexlab{a}})\citenamefont{Mizuguchi, Fujihisa, Gotoh, Suzuki,
  Usui, Kuroki, Demura, Takano, Izawa, and Miura}}]{Mizuguchi1}
\bibinfo{author}{\bibfnamefont{Y.}~\bibnamefont{Mizuguchi}},
  \bibinfo{author}{\bibfnamefont{H.}~\bibnamefont{Fujihisa}},
  \bibinfo{author}{\bibfnamefont{Y.}~\bibnamefont{Gotoh}},
  \bibinfo{author}{\bibfnamefont{K.}~\bibnamefont{Suzuki}},
  \bibinfo{author}{\bibfnamefont{H.}~\bibnamefont{Usui}},
  \bibinfo{author}{\bibfnamefont{K.}~\bibnamefont{Kuroki}},
  \bibinfo{author}{\bibfnamefont{S.}~\bibnamefont{Demura}},
  \bibinfo{author}{\bibfnamefont{Y.}~\bibnamefont{Takano}},
  \bibinfo{author}{\bibfnamefont{H.}~\bibnamefont{Izawa}}, \bibnamefont{and}
  \bibinfo{author}{\bibfnamefont{O.}~\bibnamefont{Miura}},
  \bibinfo{journal}{Phys. Rev. B} \textbf{\bibinfo{volume}{86}},
  \bibinfo{pages}{220510(R)} (\bibinfo{year}{2012}{\natexlab{a}}).

\bibitem[{\citenamefont{Li et~al.}(2012)\citenamefont{Li, Xing, and
  Huang}}]{Li}
\bibinfo{author}{\bibfnamefont{B.}~\bibnamefont{Li}},
  \bibinfo{author}{\bibfnamefont{Z.~W.} \bibnamefont{Xing}}, \bibnamefont{and}
  \bibinfo{author}{\bibfnamefont{G.~Q.} \bibnamefont{Huang}},
  \bibinfo{journal}{arXiv} p. \bibinfo{pages}{1210.1743}
  (\bibinfo{year}{2012}).

\bibitem[{\citenamefont{Jha et~al.}(2012)\citenamefont{Jha, Kumar, Singh, and
  Awana}}]{Jha}
\bibinfo{author}{\bibfnamefont{R.}~\bibnamefont{Jha}},
  \bibinfo{author}{\bibfnamefont{A.}~\bibnamefont{Kumar}},
  \bibinfo{author}{\bibfnamefont{S.~K.} \bibnamefont{Singh}}, \bibnamefont{and}
  \bibinfo{author}{\bibfnamefont{V.~P.~S.} \bibnamefont{Awana}},
  \bibinfo{journal}{arXiv} p. \bibinfo{pages}{1208.5873}
  (\bibinfo{year}{2012}).

\bibitem[{\citenamefont{Deguchi et~al.}(2012)\citenamefont{Deguchi, Mizuguchi,
  Demura, Hara, Watanabe, Denholme, Fujioka, Okazaki, Ozaki, Takeya
  et~al.}}]{Deguchi}
\bibinfo{author}{\bibfnamefont{K.}~\bibnamefont{Deguchi}},
  \bibinfo{author}{\bibfnamefont{Y.}~\bibnamefont{Mizuguchi}},
  \bibinfo{author}{\bibfnamefont{S.}~\bibnamefont{Demura}},
  \bibinfo{author}{\bibfnamefont{H.}~\bibnamefont{Hara}},
  \bibinfo{author}{\bibfnamefont{T.}~\bibnamefont{Watanabe}},
  \bibinfo{author}{\bibfnamefont{S.~J.} \bibnamefont{Denholme}},
  \bibinfo{author}{\bibfnamefont{M.}~\bibnamefont{Fujioka}},
  \bibinfo{author}{\bibfnamefont{H.}~\bibnamefont{Okazaki}},
  \bibinfo{author}{\bibfnamefont{T.}~\bibnamefont{Ozaki}},
  \bibinfo{author}{\bibfnamefont{H.}~\bibnamefont{Takeya}},
  \bibnamefont{et~al.}, \bibinfo{journal}{arXiv} p. \bibinfo{pages}{1209.3846}
  (\bibinfo{year}{2012}).

\bibitem[{\citenamefont{Kotegawa et~al.}(2012)\citenamefont{Kotegawa, Tomita,
  Tou, Izawa, Mizuguchi, Miura, Demura, Deguchi, and Takano}}]{Kotegawa}
\bibinfo{author}{\bibfnamefont{H.}~\bibnamefont{Kotegawa}},
  \bibinfo{author}{\bibfnamefont{Y.}~\bibnamefont{Tomita}},
  \bibinfo{author}{\bibfnamefont{H.}~\bibnamefont{Tou}},
  \bibinfo{author}{\bibfnamefont{H.}~\bibnamefont{Izawa}},
  \bibinfo{author}{\bibfnamefont{Y.}~\bibnamefont{Mizuguchi}},
  \bibinfo{author}{\bibfnamefont{O.}~\bibnamefont{Miura}},
  \bibinfo{author}{\bibfnamefont{S.}~\bibnamefont{Demura}},
  \bibinfo{author}{\bibfnamefont{K.}~\bibnamefont{Deguchi}}, \bibnamefont{and}
  \bibinfo{author}{\bibfnamefont{Y.}~\bibnamefont{Takano}},
  \bibinfo{journal}{J. Phys. Soc. Jpn.} \textbf{\bibinfo{volume}{81}},
  \bibinfo{pages}{103702} (\bibinfo{year}{2012}).

\bibitem[{\citenamefont{Awana et~al.}(2012)\citenamefont{Awana, Kumar, Jha,
  Kumar, Pal, Shruti, Saha, and Patnaik}}]{Awana}
\bibinfo{author}{\bibfnamefont{V.~P.~S.} \bibnamefont{Awana}},
  \bibinfo{author}{\bibfnamefont{A.}~\bibnamefont{Kumar}},
  \bibinfo{author}{\bibfnamefont{R.}~\bibnamefont{Jha}},
  \bibinfo{author}{\bibfnamefont{S.}~\bibnamefont{Kumar}},
  \bibinfo{author}{\bibfnamefont{A.}~\bibnamefont{Pal}},
  \bibinfo{author}{\bibnamefont{Shruti}},
  \bibinfo{author}{\bibfnamefont{J.}~\bibnamefont{Saha}}, \bibnamefont{and}
  \bibinfo{author}{\bibfnamefont{S.}~\bibnamefont{Patnaik}},
  \bibinfo{journal}{arXiv} p. \bibinfo{pages}{1207.6845}
  (\bibinfo{year}{2012}).

\bibitem[{\citenamefont{Demura et~al.}(2012)\citenamefont{Demura, Mizuguchi,
  Deguchi, Okazaki, Hara, Watanabe, Denholme, Fujioka, Ozaki, Fujihisa
  et~al.}}]{Demura}
\bibinfo{author}{\bibfnamefont{S.}~\bibnamefont{Demura}},
  \bibinfo{author}{\bibfnamefont{Y.}~\bibnamefont{Mizuguchi}},
  \bibinfo{author}{\bibfnamefont{K.}~\bibnamefont{Deguchi}},
  \bibinfo{author}{\bibfnamefont{H.}~\bibnamefont{Okazaki}},
  \bibinfo{author}{\bibfnamefont{H.}~\bibnamefont{Hara}},
  \bibinfo{author}{\bibfnamefont{T.}~\bibnamefont{Watanabe}},
  \bibinfo{author}{\bibfnamefont{S.~J.} \bibnamefont{Denholme}},
  \bibinfo{author}{\bibfnamefont{M.}~\bibnamefont{Fujioka}},
  \bibinfo{author}{\bibfnamefont{T.}~\bibnamefont{Ozaki}},
  \bibinfo{author}{\bibfnamefont{H.}~\bibnamefont{Fujihisa}},
  \bibnamefont{et~al.}, \bibinfo{journal}{arXiv} p. \bibinfo{pages}{1207.5248}
  (\bibinfo{year}{2012}).

\bibitem[{\citenamefont{Mizuguchi
  et~al.}(2012{\natexlab{b}})\citenamefont{Mizuguchi, Demura, Deguchi, Takano,
  Fujihisa, Gotoh, Izawa, and Miura}}]{Mizuguchi2}
\bibinfo{author}{\bibfnamefont{Y.}~\bibnamefont{Mizuguchi}},
  \bibinfo{author}{\bibfnamefont{S.}~\bibnamefont{Demura}},
  \bibinfo{author}{\bibfnamefont{K.}~\bibnamefont{Deguchi}},
  \bibinfo{author}{\bibfnamefont{Y.}~\bibnamefont{Takano}},
  \bibinfo{author}{\bibfnamefont{H.}~\bibnamefont{Fujihisa}},
  \bibinfo{author}{\bibfnamefont{Y.}~\bibnamefont{Gotoh}},
  \bibinfo{author}{\bibfnamefont{H.}~\bibnamefont{Izawa}}, \bibnamefont{and}
  \bibinfo{author}{\bibfnamefont{O.}~\bibnamefont{Miura}}, \bibinfo{journal}{J.
  Phys. Soc. Jap.} \textbf{\bibinfo{volume}{81}}, \bibinfo{pages}{114725}
  (\bibinfo{year}{2012}{\natexlab{b}}).

\bibitem[{\citenamefont{Xing et~al.}(2012)\citenamefont{Xing, Li, Ding, Yang,
  and Wen}}]{Xing}
\bibinfo{author}{\bibfnamefont{J.}~\bibnamefont{Xing}},
  \bibinfo{author}{\bibfnamefont{S.}~\bibnamefont{Li}},
  \bibinfo{author}{\bibfnamefont{X.}~\bibnamefont{Ding}},
  \bibinfo{author}{\bibfnamefont{H.}~\bibnamefont{Yang}}, \bibnamefont{and}
  \bibinfo{author}{\bibfnamefont{H.-H.} \bibnamefont{Wen}},
  \bibinfo{journal}{arxiv} p. \bibinfo{pages}{1208.5000}
  (\bibinfo{year}{2012}).

\bibitem[{\citenamefont{Yazici et~al.}(2012)\citenamefont{Yazici, Huang, White,
  Chang, Friedman, and Maple}}]{Yazici}
\bibinfo{author}{\bibfnamefont{D.}~\bibnamefont{Yazici}},
  \bibinfo{author}{\bibfnamefont{K.}~\bibnamefont{Huang}},
  \bibinfo{author}{\bibfnamefont{B.~D.} \bibnamefont{White}},
  \bibinfo{author}{\bibfnamefont{A.~H.} \bibnamefont{Chang}},
  \bibinfo{author}{\bibfnamefont{A.~J.} \bibnamefont{Friedman}},
  \bibnamefont{and} \bibinfo{author}{\bibfnamefont{M.~B.} \bibnamefont{Maple}},
  \bibinfo{journal}{Philos. Mag.} \textbf{\bibinfo{volume}{93}},
  \bibinfo{pages}{673} (\bibinfo{year}{2012}).

\bibitem[{\citenamefont{Zhano et~al.}(2008)\citenamefont{Zhano, Huang, Cruz,
  Li, Lynn, Chen, Green, Chen, Li, Li et~al.}}]{Zhano}
\bibinfo{author}{\bibfnamefont{J.}~\bibnamefont{Zhano}},
  \bibinfo{author}{\bibfnamefont{Q.}~\bibnamefont{Huang}},
  \bibinfo{author}{\bibfnamefont{C.~D.~L.} \bibnamefont{Cruz}},
  \bibinfo{author}{\bibfnamefont{C.}~\bibnamefont{Li}},
  \bibinfo{author}{\bibfnamefont{J.~W.} \bibnamefont{Lynn}},
  \bibinfo{author}{\bibfnamefont{Y.}~\bibnamefont{Chen}},
  \bibinfo{author}{\bibfnamefont{M.~A.} \bibnamefont{Green}},
  \bibinfo{author}{\bibfnamefont{G.~F.} \bibnamefont{Chen}},
  \bibinfo{author}{\bibfnamefont{G.}~\bibnamefont{Li}},
  \bibinfo{author}{\bibfnamefont{Z.}~\bibnamefont{Li}}, \bibnamefont{et~al.},
  \bibinfo{journal}{Nature Materials} \textbf{\bibinfo{volume}{7}},
  \bibinfo{pages}{953} (\bibinfo{year}{2008}).

\bibitem[{\citenamefont{Alff et~al.}(2003)\citenamefont{Alff, Krockenberger,
  Welter, Gross, Manske, and Naito}}]{Alff}
\bibinfo{author}{\bibfnamefont{L.}~\bibnamefont{Alff}},
  \bibinfo{author}{\bibfnamefont{Y.}~\bibnamefont{Krockenberger}},
  \bibinfo{author}{\bibfnamefont{B.}~\bibnamefont{Welter}},
  \bibinfo{author}{\bibfnamefont{R.}~\bibnamefont{Gross}},
  \bibinfo{author}{\bibfnamefont{D.}~\bibnamefont{Manske}}, \bibnamefont{and}
  \bibinfo{author}{\bibfnamefont{M.}~\bibnamefont{Naito}},
  \bibinfo{journal}{Nature (London)} \textbf{\bibinfo{volume}{422}},
  \bibinfo{pages}{698} (\bibinfo{year}{2003}).

\bibitem[{\citenamefont{Paglione and Greene}(2010)}]{Paglione}
\bibinfo{author}{\bibfnamefont{J.}~\bibnamefont{Paglione}} \bibnamefont{and}
  \bibinfo{author}{\bibfnamefont{R.~L.} \bibnamefont{Greene}},
  \bibinfo{journal}{Nature Physics} \textbf{\bibinfo{volume}{6}},
  \bibinfo{pages}{645} (\bibinfo{year}{2010}).

\bibitem[{\citenamefont{Johnston}(2010)}]{Johnston}
\bibinfo{author}{\bibfnamefont{D.~C.} \bibnamefont{Johnston}},
  \bibinfo{journal}{Advances in Physics} \textbf{\bibinfo{volume}{59}}
  (\bibinfo{year}{2010}).

\bibitem[{\citenamefont{Mazin}(2010)}]{Mazin}
\bibinfo{author}{\bibfnamefont{I.~I.} \bibnamefont{Mazin}},
  \bibinfo{journal}{Nature} \textbf{\bibinfo{volume}{464}}
  (\bibinfo{year}{2010}).

\bibitem[{\citenamefont{Okada et~al.}(2008)\citenamefont{Okada, Igawa,
  Takahashi, Kamihara, Hirano, Hosono, Matsubayashi, and
  Uwatoko}}]{okada_2008_1}
\bibinfo{author}{\bibfnamefont{H.}~\bibnamefont{Okada}},
  \bibinfo{author}{\bibfnamefont{K.}~\bibnamefont{Igawa}},
  \bibinfo{author}{\bibfnamefont{H.}~\bibnamefont{Takahashi}},
  \bibinfo{author}{\bibfnamefont{Y.}~\bibnamefont{Kamihara}},
  \bibinfo{author}{\bibfnamefont{M.}~\bibnamefont{Hirano}},
  \bibinfo{author}{\bibfnamefont{H.}~\bibnamefont{Hosono}},
  \bibinfo{author}{\bibfnamefont{K.}~\bibnamefont{Matsubayashi}},
  \bibnamefont{and} \bibinfo{author}{\bibfnamefont{Y.}~\bibnamefont{Uwatoko}},
  \bibinfo{journal}{Journal of the Physical Society of Japan}
  \textbf{\bibinfo{volume}{77}}, \bibinfo{pages}{113712}
  (\bibinfo{year}{2008}).

\bibitem[{\citenamefont{Hamlin et~al.}(2008)\citenamefont{Hamlin, Baumbach,
  Zocco, Sayles, and Maple}}]{Hamlin}
\bibinfo{author}{\bibfnamefont{J.~J.} \bibnamefont{Hamlin}},
  \bibinfo{author}{\bibfnamefont{R.~E.} \bibnamefont{Baumbach}},
  \bibinfo{author}{\bibfnamefont{D.~A.} \bibnamefont{Zocco}},
  \bibinfo{author}{\bibfnamefont{T.~A.} \bibnamefont{Sayles}},
  \bibnamefont{and} \bibinfo{author}{\bibfnamefont{M.~B.} \bibnamefont{Maple}},
  \bibinfo{journal}{Journal of Physics: Condensed Matter}
  \textbf{\bibinfo{volume}{20}}, \bibinfo{pages}{365220}
  (\bibinfo{year}{2008}).

\bibitem[{\citenamefont{Kamihara
  et~al.}(2008{\natexlab{a}})\citenamefont{Kamihara, Hiramatsu, Hirano,
  Kawamura, Yanagi, Kamiya, and Hosono}}]{Kamihara}
\bibinfo{author}{\bibfnamefont{Y.}~\bibnamefont{Kamihara}},
  \bibinfo{author}{\bibfnamefont{H.}~\bibnamefont{Hiramatsu}},
  \bibinfo{author}{\bibfnamefont{M.}~\bibnamefont{Hirano}},
  \bibinfo{author}{\bibfnamefont{R.}~\bibnamefont{Kawamura}},
  \bibinfo{author}{\bibfnamefont{H.}~\bibnamefont{Yanagi}},
  \bibinfo{author}{\bibfnamefont{T.}~\bibnamefont{Kamiya}}, \bibnamefont{and}
  \bibinfo{author}{\bibfnamefont{H.}~\bibnamefont{Hosono}},
  \bibinfo{journal}{Journal of Physics: Condensed Matter}
  \textbf{\bibinfo{volume}{20}}, \bibinfo{pages}{365220}
  (\bibinfo{year}{2008}{\natexlab{a}}).

\bibitem[{\citenamefont{Tegel et~al.}(2008)\citenamefont{Tegel, Schellenberg,
  P{\"o}ttgen, and Johrendt}}]{tegel_2008_1}
\bibinfo{author}{\bibfnamefont{M.}~\bibnamefont{Tegel}},
  \bibinfo{author}{\bibfnamefont{I.}~\bibnamefont{Schellenberg}},
  \bibinfo{author}{\bibfnamefont{R.}~\bibnamefont{P{\"o}ttgen}},
  \bibnamefont{and} \bibinfo{author}{\bibfnamefont{D.}~\bibnamefont{Johrendt}},
  \bibinfo{journal}{Z. Naturforsch. B - Chem. Sci.}
  \textbf{\bibinfo{volume}{63}}, \bibinfo{pages}{1057} (\bibinfo{year}{2008}).

\bibitem[{\citenamefont{Kamihara
  et~al.}(2008{\natexlab{b}})\citenamefont{Kamihara, Watanabe, Hirano, and
  Hosono}}]{kamihara_2008_1}
\bibinfo{author}{\bibfnamefont{Y.}~\bibnamefont{Kamihara}},
  \bibinfo{author}{\bibfnamefont{T.}~\bibnamefont{Watanabe}},
  \bibinfo{author}{\bibfnamefont{M.}~\bibnamefont{Hirano}}, \bibnamefont{and}
  \bibinfo{author}{\bibfnamefont{H.}~\bibnamefont{Hosono}},
  \bibinfo{journal}{Journal of the American Chemical Society}
  \textbf{\bibinfo{volume}{130}}, \bibinfo{pages}{3296}
  (\bibinfo{year}{2008}{\natexlab{b}}).

\bibitem[{\citenamefont{Chen et~al.}(2008{\natexlab{a}})\citenamefont{Chen, Li,
  Wu, Li, Hu, Dong, Zheng, Luo, and Wang}}]{chen_2008_1}
\bibinfo{author}{\bibfnamefont{G.~F.} \bibnamefont{Chen}},
  \bibinfo{author}{\bibfnamefont{Z.}~\bibnamefont{Li}},
  \bibinfo{author}{\bibfnamefont{D.}~\bibnamefont{Wu}},
  \bibinfo{author}{\bibfnamefont{G.}~\bibnamefont{Li}},
  \bibinfo{author}{\bibfnamefont{Z.}~\bibnamefont{Hu}},
  \bibinfo{author}{\bibfnamefont{J.}~\bibnamefont{Dong}},
  \bibinfo{author}{\bibfnamefont{P.}~\bibnamefont{Zheng}},
  \bibinfo{author}{\bibfnamefont{J.~L.} \bibnamefont{Luo}}, \bibnamefont{and}
  \bibinfo{author}{\bibfnamefont{N.~L.} \bibnamefont{Wang}},
  \bibinfo{journal}{Physical Review Letters} \textbf{\bibinfo{volume}{100}},
  \bibinfo{pages}{247002} (\bibinfo{year}{2008}{\natexlab{a}}).

\bibitem[{\citenamefont{Ren et~al.}(2008{\natexlab{a}})\citenamefont{Ren, Lu,
  Yang, Yi, Shen, Li, Che, Dong, Sun, Zhou et~al.}}]{ren_2008_1}
\bibinfo{author}{\bibfnamefont{Z.~A.} \bibnamefont{Ren}},
  \bibinfo{author}{\bibfnamefont{W.}~\bibnamefont{Lu}},
  \bibinfo{author}{\bibfnamefont{J.}~\bibnamefont{Yang}},
  \bibinfo{author}{\bibfnamefont{W.}~\bibnamefont{Yi}},
  \bibinfo{author}{\bibfnamefont{X.~L.} \bibnamefont{Shen}},
  \bibinfo{author}{\bibfnamefont{Z.~C.} \bibnamefont{Li}},
  \bibinfo{author}{\bibfnamefont{G.~C.} \bibnamefont{Che}},
  \bibinfo{author}{\bibfnamefont{X.~L.} \bibnamefont{Dong}},
  \bibinfo{author}{\bibfnamefont{L.~L.} \bibnamefont{Sun}},
  \bibinfo{author}{\bibfnamefont{F.}~\bibnamefont{Zhou}}, \bibnamefont{et~al.},
  \bibinfo{journal}{Chinese Physics Letters} \textbf{\bibinfo{volume}{25}},
  \bibinfo{pages}{2215} (\bibinfo{year}{2008}{\natexlab{a}}).

\bibitem[{\citenamefont{Ren et~al.}(2008{\natexlab{b}})\citenamefont{Ren, Yang,
  Lu, Yi, Che, Dong, Sun, and Zhao}}]{ren_2008_3}
\bibinfo{author}{\bibfnamefont{Z.~A.} \bibnamefont{Ren}},
  \bibinfo{author}{\bibfnamefont{J.}~\bibnamefont{Yang}},
  \bibinfo{author}{\bibfnamefont{W.}~\bibnamefont{Lu}},
  \bibinfo{author}{\bibfnamefont{W.}~\bibnamefont{Yi}},
  \bibinfo{author}{\bibfnamefont{G.~C.} \bibnamefont{Che}},
  \bibinfo{author}{\bibfnamefont{X.~L.} \bibnamefont{Dong}},
  \bibinfo{author}{\bibfnamefont{L.~L.} \bibnamefont{Sun}}, \bibnamefont{and}
  \bibinfo{author}{\bibfnamefont{Z.-X.} \bibnamefont{Zhao}},
  \bibinfo{journal}{Materials Research Innovations}
  \textbf{\bibinfo{volume}{12}}, \bibinfo{pages}{105}
  (\bibinfo{year}{2008}{\natexlab{b}}).

\bibitem[{\citenamefont{Ren et~al.}(2008{\natexlab{c}})\citenamefont{Ren, Yang,
  Lu, Yi, Shen, Li, Che, Dong, Sun, Zhou et~al.}}]{ren_2008_2}
\bibinfo{author}{\bibfnamefont{Z.-A.} \bibnamefont{Ren}},
  \bibinfo{author}{\bibfnamefont{J.}~\bibnamefont{Yang}},
  \bibinfo{author}{\bibfnamefont{W.}~\bibnamefont{Lu}},
  \bibinfo{author}{\bibfnamefont{W.}~\bibnamefont{Yi}},
  \bibinfo{author}{\bibfnamefont{X.~L.} \bibnamefont{Shen}},
  \bibinfo{author}{\bibfnamefont{Z.~C.} \bibnamefont{Li}},
  \bibinfo{author}{\bibfnamefont{G.~C.} \bibnamefont{Che}},
  \bibinfo{author}{\bibfnamefont{X.~L.} \bibnamefont{Dong}},
  \bibinfo{author}{\bibfnamefont{L.~L.} \bibnamefont{Sun}},
  \bibinfo{author}{\bibfnamefont{F.}~\bibnamefont{Zhou}}, \bibnamefont{et~al.},
  \bibinfo{journal}{Europhysics Letters} \textbf{\bibinfo{volume}{82}},
  \bibinfo{pages}{57002} (\bibinfo{year}{2008}{\natexlab{c}}).

\bibitem[{\citenamefont{Chen et~al.}(2008{\natexlab{b}})\citenamefont{Chen, Wu,
  Liu, Chen, and Fang}}]{fang_2008_2}
\bibinfo{author}{\bibfnamefont{X.~H.} \bibnamefont{Chen}},
  \bibinfo{author}{\bibfnamefont{T.}~\bibnamefont{Wu}},
  \bibinfo{author}{\bibfnamefont{R.~H.} \bibnamefont{Liu}},
  \bibinfo{author}{\bibfnamefont{H.}~\bibnamefont{Chen}}, \bibnamefont{and}
  \bibinfo{author}{\bibfnamefont{D.~F.} \bibnamefont{Fang}},
  \bibinfo{journal}{Nature} \textbf{\bibinfo{volume}{453}},
  \bibinfo{pages}{761} (\bibinfo{year}{2008}{\natexlab{b}}).

\bibitem[{\citenamefont{Sefat et~al.}(2008)\citenamefont{Sefat, Huq, McGuire,
  Jin, Sales, Mandrus, Cranswick, Stephens, and Stone}}]{sales_2008_2}
\bibinfo{author}{\bibfnamefont{A.~S.} \bibnamefont{Sefat}},
  \bibinfo{author}{\bibfnamefont{A.}~\bibnamefont{Huq}},
  \bibinfo{author}{\bibfnamefont{M.~A.} \bibnamefont{McGuire}},
  \bibinfo{author}{\bibfnamefont{R.}~\bibnamefont{Jin}},
  \bibinfo{author}{\bibfnamefont{B.~C.} \bibnamefont{Sales}},
  \bibinfo{author}{\bibfnamefont{D.}~\bibnamefont{Mandrus}},
  \bibinfo{author}{\bibfnamefont{L.~M.~D.} \bibnamefont{Cranswick}},
  \bibinfo{author}{\bibfnamefont{P.~W.} \bibnamefont{Stephens}},
  \bibnamefont{and} \bibinfo{author}{\bibfnamefont{K.~H.} \bibnamefont{Stone}},
  \bibinfo{journal}{Physical Review B} \textbf{\bibinfo{volume}{78}},
  \bibinfo{pages}{104505} (\bibinfo{year}{2008}).

\bibitem[{\citenamefont{Wen et~al.}(2008)\citenamefont{Wen, Mu, Fang, Yang, and
  Zhu}}]{zhu_2008_1}
\bibinfo{author}{\bibfnamefont{H.-H.} \bibnamefont{Wen}},
  \bibinfo{author}{\bibfnamefont{G.}~\bibnamefont{Mu}},
  \bibinfo{author}{\bibfnamefont{L.}~\bibnamefont{Fang}},
  \bibinfo{author}{\bibfnamefont{H.}~\bibnamefont{Yang}}, \bibnamefont{and}
  \bibinfo{author}{\bibfnamefont{Y.}~\bibnamefont{Zhu}},
  \bibinfo{journal}{Europhysics Letters} \textbf{\bibinfo{volume}{82}},
  \bibinfo{pages}{17009} (\bibinfo{year}{2008}).

\bibitem[{\citenamefont{Wang et~al.}(2008)\citenamefont{Wang, Li, Chi, Zhu,
  Ren, Li, Wang, Lin, Luo, Jiang et~al.}}]{ren_2008_5}
\bibinfo{author}{\bibfnamefont{C.}~\bibnamefont{Wang}},
  \bibinfo{author}{\bibfnamefont{L.}~\bibnamefont{Li}},
  \bibinfo{author}{\bibfnamefont{S.}~\bibnamefont{Chi}},
  \bibinfo{author}{\bibfnamefont{Z.}~\bibnamefont{Zhu}},
  \bibinfo{author}{\bibfnamefont{Z.}~\bibnamefont{Ren}},
  \bibinfo{author}{\bibfnamefont{Y.}~\bibnamefont{Li}},
  \bibinfo{author}{\bibfnamefont{Y.}~\bibnamefont{Wang}},
  \bibinfo{author}{\bibfnamefont{X.}~\bibnamefont{Lin}},
  \bibinfo{author}{\bibfnamefont{Y.}~\bibnamefont{Luo}},
  \bibinfo{author}{\bibfnamefont{S.}~\bibnamefont{Jiang}},
  \bibnamefont{et~al.}, \bibinfo{journal}{Europhysics Letters}
  \textbf{\bibinfo{volume}{83}}, \bibinfo{pages}{67006} (\bibinfo{year}{2008}).

\bibitem[{\citenamefont{Ren et~al.}(2008{\natexlab{d}})\citenamefont{Ren, Che,
  Dong, Yang, Lu, Yi, Shen, Li, Sun, Zhou et~al.}}]{ren_2008_6}
\bibinfo{author}{\bibfnamefont{Z.-A.} \bibnamefont{Ren}},
  \bibinfo{author}{\bibfnamefont{G.-C.} \bibnamefont{Che}},
  \bibinfo{author}{\bibfnamefont{X.-L.} \bibnamefont{Dong}},
  \bibinfo{author}{\bibfnamefont{J.}~\bibnamefont{Yang}},
  \bibinfo{author}{\bibfnamefont{W.}~\bibnamefont{Lu}},
  \bibinfo{author}{\bibfnamefont{W.}~\bibnamefont{Yi}},
  \bibinfo{author}{\bibfnamefont{X.-L.} \bibnamefont{Shen}},
  \bibinfo{author}{\bibfnamefont{Z.-C.} \bibnamefont{Li}},
  \bibinfo{author}{\bibfnamefont{L.-L.} \bibnamefont{Sun}},
  \bibinfo{author}{\bibfnamefont{F.}~\bibnamefont{Zhou}}, \bibnamefont{et~al.},
  \bibinfo{journal}{Europhysics Letters} \textbf{\bibinfo{volume}{83}},
  \bibinfo{pages}{17002} (\bibinfo{year}{2008}{\natexlab{d}}).

\bibitem[{\citenamefont{Yang et~al.}(2008)\citenamefont{Yang, Li, Lu, Yi, Shen,
  Ren, Che, Dong, Sun, Zhou et~al.}}]{yang_2008_1}
\bibinfo{author}{\bibfnamefont{J.}~\bibnamefont{Yang}},
  \bibinfo{author}{\bibfnamefont{Z.-C.} \bibnamefont{Li}},
  \bibinfo{author}{\bibfnamefont{W.}~\bibnamefont{Lu}},
  \bibinfo{author}{\bibfnamefont{W.}~\bibnamefont{Yi}},
  \bibinfo{author}{\bibfnamefont{X.-L.} \bibnamefont{Shen}},
  \bibinfo{author}{\bibfnamefont{Z.-A.} \bibnamefont{Ren}},
  \bibinfo{author}{\bibfnamefont{G.-C.} \bibnamefont{Che}},
  \bibinfo{author}{\bibfnamefont{X.-L.} \bibnamefont{Dong}},
  \bibinfo{author}{\bibfnamefont{L.-L.} \bibnamefont{Sun}},
  \bibinfo{author}{\bibfnamefont{F.}~\bibnamefont{Zhou}}, \bibnamefont{et~al.},
  \bibinfo{journal}{Superconductor Science and Technology}
  \textbf{\bibinfo{volume}{21}}, \bibinfo{pages}{082001}
  (\bibinfo{year}{2008}).

\bibitem[{\citenamefont{Yildirim}(2012)}]{Yildirim}
\bibinfo{author}{\bibfnamefont{T.}~\bibnamefont{Yildirim}},
  \bibinfo{journal}{arXiv} p. \bibinfo{pages}{1210.2418}
  (\bibinfo{year}{2012}).

\bibitem[{\citenamefont{Lee et~al.}(2012)\citenamefont{Lee, Stone, Huq,
  Yildirim, Ehlers, Mizuguchi, Miura, Takano, Deguchi, Demura et~al.}}]{Lee}
\bibinfo{author}{\bibfnamefont{J.}~\bibnamefont{Lee}},
  \bibinfo{author}{\bibfnamefont{M.~B.} \bibnamefont{Stone}},
  \bibinfo{author}{\bibfnamefont{A.}~\bibnamefont{Huq}},
  \bibinfo{author}{\bibfnamefont{T.}~\bibnamefont{Yildirim}},
  \bibinfo{author}{\bibfnamefont{G.}~\bibnamefont{Ehlers}},
  \bibinfo{author}{\bibfnamefont{Y.}~\bibnamefont{Mizuguchi}},
  \bibinfo{author}{\bibfnamefont{O.}~\bibnamefont{Miura}},
  \bibinfo{author}{\bibfnamefont{Y.}~\bibnamefont{Takano}},
  \bibinfo{author}{\bibfnamefont{K.}~\bibnamefont{Deguchi}},
  \bibinfo{author}{\bibfnamefont{S.}~\bibnamefont{Demura}},
  \bibnamefont{et~al.}, \bibinfo{journal}{arXiv} p. \bibinfo{pages}{1212.4811}
  (\bibinfo{year}{2012}).

\bibitem[{\citenamefont{Shein and Ivanovskii}(2012)}]{Shein}
\bibinfo{author}{\bibfnamefont{I.~R.} \bibnamefont{Shein}} \bibnamefont{and}
  \bibinfo{author}{\bibfnamefont{A.~L.} \bibnamefont{Ivanovskii}},
  \bibinfo{journal}{arXiv} p. \bibinfo{pages}{1211.3818}
  (\bibinfo{year}{2012}).

\bibitem[{\citenamefont{Wan et~al.}(2012)\citenamefont{Wan, Ding, Savrasov, and
  Duan}}]{Wan}
\bibinfo{author}{\bibfnamefont{X.}~\bibnamefont{Wan}},
  \bibinfo{author}{\bibfnamefont{H.-C.} \bibnamefont{Ding}},
  \bibinfo{author}{\bibfnamefont{S.~Y.} \bibnamefont{Savrasov}},
  \bibnamefont{and} \bibinfo{author}{\bibfnamefont{C.-G.} \bibnamefont{Duan}},
  \bibinfo{journal}{arXiv} p. \bibinfo{pages}{1208.1807}
  (\bibinfo{year}{2012}).

\bibitem[{\citenamefont{Rietveld}(1969)}]{Rietveld}
\bibinfo{author}{\bibfnamefont{H.~M.} \bibnamefont{Rietveld}},
  \bibinfo{journal}{J. Appl. Cryst.} \textbf{\bibinfo{volume}{2}},
  \bibinfo{pages}{65} (\bibinfo{year}{1969}).

\bibitem[{\citenamefont{Larson and Von~Dreele}(1994)}]{LARSON01}
\bibinfo{author}{\bibfnamefont{A.~C.} \bibnamefont{Larson}} \bibnamefont{and}
  \bibinfo{author}{\bibfnamefont{R.~B.} \bibnamefont{Von~Dreele}},
  \emph{\bibinfo{title}{General Structure Analysis System (GSAS)}},
  \bibinfo{organization}{Los Alamos National Laboratory Report}
  (\bibinfo{year}{1994}).

\bibitem[{\citenamefont{Toby}(2001)}]{expgui}
\bibinfo{author}{\bibfnamefont{B.~H.} \bibnamefont{Toby}}, \bibinfo{journal}{J.
  Appl. Cryst.} \textbf{\bibinfo{volume}{34}}, \bibinfo{pages}{210}
  (\bibinfo{year}{2001}).

\bibitem[{\citenamefont{Safar et~al.}(1992)\citenamefont{Safar, Gammel, Huse,
  Bishop, Rice, and Ginsberg}}]{Safar}
\bibinfo{author}{\bibfnamefont{H.}~\bibnamefont{Safar}},
  \bibinfo{author}{\bibfnamefont{P.~L.} \bibnamefont{Gammel}},
  \bibinfo{author}{\bibfnamefont{D.~A.} \bibnamefont{Huse}},
  \bibinfo{author}{\bibfnamefont{D.~J.} \bibnamefont{Bishop}},
  \bibinfo{author}{\bibfnamefont{J.~P.} \bibnamefont{Rice}}, \bibnamefont{and}
  \bibinfo{author}{\bibfnamefont{D.~M.} \bibnamefont{Ginsberg}},
  \bibinfo{journal}{Phys. Rev. Lett.} \textbf{\bibinfo{volume}{69}},
  \bibinfo{pages}{824} (\bibinfo{year}{1992}).

\bibitem[{\citenamefont{Werthamer et~al.}(1966)\citenamefont{Werthamer,
  Helfand, and Hohenberg}}]{Werthamer}
\bibinfo{author}{\bibfnamefont{N.~R.} \bibnamefont{Werthamer}},
  \bibinfo{author}{\bibfnamefont{E.}~\bibnamefont{Helfand}}, \bibnamefont{and}
  \bibinfo{author}{\bibfnamefont{P.~C.} \bibnamefont{Hohenberg}},
  \bibinfo{journal}{Physical Review} \textbf{\bibinfo{volume}{147}},
  \bibinfo{pages}{1} (\bibinfo{year}{1966}).

\bibitem[{\citenamefont{Lin et~al.}(2013)\citenamefont{Lin, Ni, Chen, Xu, Yang,
  Dai, Li, Yang, Luo, Tao et~al.}}]{Lin}
\bibinfo{author}{\bibfnamefont{X.}~\bibnamefont{Lin}},
  \bibinfo{author}{\bibfnamefont{X.}~\bibnamefont{Ni}},
  \bibinfo{author}{\bibfnamefont{B.}~\bibnamefont{Chen}},
  \bibinfo{author}{\bibfnamefont{X.}~\bibnamefont{Xu}},
  \bibinfo{author}{\bibfnamefont{X.}~\bibnamefont{Yang}},
  \bibinfo{author}{\bibfnamefont{J.}~\bibnamefont{Dai}},
  \bibinfo{author}{\bibfnamefont{Y.}~\bibnamefont{Li}},
  \bibinfo{author}{\bibfnamefont{X.}~\bibnamefont{Yang}},
  \bibinfo{author}{\bibfnamefont{Y.}~\bibnamefont{Luo}},
  \bibinfo{author}{\bibfnamefont{Q.}~\bibnamefont{Tao}}, \bibnamefont{et~al.},
  \bibinfo{journal}{Phys. Rev. B} \textbf{\bibinfo{volume}{87}},
  \bibinfo{pages}{020504(R)} (\bibinfo{year}{2013}).

\end{thebibliography}

\end{document}